# Accretion discs with accreting coronae in AGN.
# I. Solutions in hydrostatic equilibrium

Piotr T. Życki[1], Suzy Collin-Souffrin[2,3] and Bożena Czerny[1]
[1] *Nicolaus Copernicus Astronomical Center, Bartycka 18, 00-716 Warsaw, Poland*
[2] *DAEC, Observatoire de Paris – Meudon, Place Jansen, F-92195 Meudon, France*
[3] *Institut d'Astrophysique, 98bis Bld Arago, F-75014 Paris, France*

**ABSTRACT**

We discuss a model for the central region of radio-quiet AGN in which the coronal accretion is the source of energy for X-ray emission. We consider two–layer geometry and we construct solutions taking into account radiative interaction and pressure balance between the disc and the corona. We define three models of the disc–corona transition region. We use two descriptions of the angular momentum transport: either the angular momentum is transported locally both in the disc and in the corona or all the angular momentum is transported within the disc only. We employ the $\alpha$–description of viscosity within the two layers but we discuss also models parameterized by a fraction of energy dissipated in the corona. Both layers are treated using vertical averages but with a clear distinction between the values of the pressure and other relevant functions at the disc surface and the mean values inside the disc.

Both descriptions of the angular momentum transport lead to similar conclusions about the fraction of energy liberated in the disc although they predict strongly different properties of the disc interiors.

The coupled disc/corona system has one family of solutions at low accretion rates, two families of solutions at intermediate accretion rates and no solution for high accretion rate. Of the two families of solutions, the first one has weakly dissipating coronae and the second one has coronae which dominate energetically the system. The two solutions merge at sub-Eddington accretion rate because a corona in hydrostatic equilibrium can dissipate only a limited amount of energy, which depends on the viscosity parameter.

At higher accretion rates an outflow from the corona in the vertical direction should be taken into account.

## 1 INTRODUCTION

The overall X-ray spectrum of a typical Seyfert galaxy consists of the primary featureless component (a power law with energy index $\sim 0.8-1.0$) and a component being a reflection of the primary emission from cold gas (Pounds et al. 1990 and subsequent papers; see Nandra & Pounds 1994 for recent update). Approximately half of the primary radiation is seen unprocessed and half of it is reflected.

Recent data on X-ray and $\gamma$-ray spectrum of Seyfert galaxy NGC 4151 from OSSE instruments on board of Compton GRO suggest that the high frequency emission is predominantly of thermal character, since the high energy cut-off is observed at only $\sim 120$ keV (Maisack et al. 1993) although in the second source measured (IC 4329a) the high frequency cut-off of the primary component was significantly higher ($240$ keV $< E_{\rm C} < 900$ keV; Madejski et al. 1995). This thermal interpretation of the X-ray emission is also strongly supported (although not proved) by the shape of the cosmic X-ray background, which is thought produced by the overall population of active galactic nuclei (see e.g. Zdziarski, Życki & Krolik 1993).

Although it is widely believed that accretion onto supermassive black hole is the ultimate source of power in AGN we do not know how this process proceeds; in particular, we do not know where and how the X-ray emission is produced. Therefore, a number of scenarios of the generation of X-ray emission from accreting black holes have been suggested so far: emission from optically thin inner part of accretion disc (Shapiro, Lightman & Eardley 1976), emission from shocks in the region of jet formation (Henri & Pelletier 1991), emission from hot plasma heated by accreting cold blobs (Sivron & Tsuruta 1993) and emission from accretion disc corona (Liang & Price 1977).

Coronal emission is particularly attractive as it is the only scenario which clearly predicts that $\sim$ half of the primary (coronal) radiation is intercepted by the cool accretion disc. Therefore in this paper we concentrate on this general scenario.

Most of the previous results on the accretion disc coro-nae concern three cases: (i) corona heated by an external unspecified X-ray source (Begelman, McKee & Shields 1983, Ostriker, McKee & Klein 1991) (ii) corona heated by energy flux transported from the disc, e.g. through acoustic or magnetohydrodynamical waves (Liang & Price 1977; Bisnovatyi-Kogan & Blinnikov 1977; Galeev, Rosner & Vaiana 1979) (iii) weak corona heated by viscous forces (Shaviv & Wehrse 1986).

The first approach is interesting for the outer parts of an accretion disc but it does not solve the problem of the origin of the X-ray emission. The second approach is not much promising as only a small fraction of energy can generally be transported to the corona (but see, e.g., Burn & Kuperus 1988) whilst in Seyfert galaxies a large part of the energy seems to originate primarily in the form of X-rays (Czerny & Życki 1994; this fraction seems to be smaller in bright AGN – quasars – as indicated by an average steeper mean $\alpha_{o/x}$ spectral slope and weaker reflected component, e.g. Williams et al. 1992). The third type of models was discussed for gas dominated discs (Shaviv & Wehrse 1986, Czerny & King 1989) or were based on the assumption that the energy generation is proportional to the gas density and not to the pressure (Shimura & Takahara 1993). These assumptions prevented the formation of a strongly dissipative corona.

The fourth option, i.e. generation of all the energy within the corona was suggested by Paczyński (1978). He assumed that half of the released energy is emitted in X and $\gamma$ band and that the other half is intercepted by an isothermal disc. The structure of the corona was not discussed in the paper; instead, an assumption was made that the disc itself is marginally self–gravitating.

Liang & Thompson (1979) actually calculated the structure of the corona. They assumed that only a fraction $f$ of the energy is dissipated in the corona; the corona is a two–temperature plasma (with Coulomb coupling between electrons and ions, as in Shapiro, Lightman & Eardley 1976), it is uniform and satisfies the hydrostatic equilibrium. This model postulates the equipartition of the radiative and conductive fluxes, which then uniquely determines the solution. Such a corona has a height comparable to the disc itself and the density is low enough to keep it optically thin. A more general approach to the geometry of the cold and the hot gas was discussed by Wandel & Liang (1991).

Both assumptions (hydrostatic equilibrium and marginal conductivity) were relaxed by Haardt & Maraschi (1991; hereafter HM91, see also Haardt & Maraschi 1993; hereafter HM93). Instead, they parameterize the solution simply by the fraction of energy dissipated in the corona, $f$, and its optical depth, $\tau$. Such an approach has the clear advantage of being very general but it does not allow to predict the influence of the existence of the corona on the disc structure, and vice-versa. Two recent papers addressed this issue.

The requirement of hydrostatic equilibrium was reintroduced by Kusunose & Mineshige (1994). They parameterize their solution with the fraction of energy dissipated in the corona, $f$, and calculate the vertically averaged structure of the disc and the corona under the assumption that the mean pressure is the same in both. However, the mean pressure in the corona is significantly lower than the mean pressure in the disc (Svenssson & Zdziarski 1994; hereafter SZ94, Czerny 1994). This modification allowed for broader range of solutions (SZ94) than found by Kusunose & Mineshige (1994).

Nakamura & Osaki (1993; hereafter NO93) replaced the arbitrary parameter $f$ by assuming that the dissipation in the corona can be modeled in the way usually adopted for the disc, i.e. that the flux is generated by viscosity which scales with the pressure. Also in that paper the hydrostatic equilibrium was described assuming the same pressure in the disc and in the corona.

In the present paper we rediscuss the existence and the structure of strong coronae within a new frame. We assume that coronal accretion itself is the source of energy of the corona, whereas the energy released within the main body of the disc is subsequently transported to the corona. A similar scenario has been recently suggested by Meyer & Meyer-Hoffmeister (1994) in the context of cataclysmic variable VW Hyi.

We describe the coronal accretion adopting the popular $\alpha$–viscosity description. We consider two cases of the angular momentum transfer. In the first model (local viscosity model – LVM) we assume that the viscosity operates locally both in the disc and in the corona and both the disc and the corona transport the angular momentum. In the second model (global viscosity model – GVM) we assume there exists a global coupling between the two layers so the entire angular momentum is transported by the disc. We also consider models for the spontaneous division of the flow between the cold and hot parts. Therefore our models *predict* the ratio of the energy generated in the disc to that generated in the corona. This ratio is, in general, a function of radius so there is some exchange of angular momentum and mass between the disc and the corona.

Our models are parametrized by the viscosity coefficients $\alpha$ in the disc and in the corona, the value of the central mass and the accretion rate. We compare our results with that of models parameterized by an arbitrarily assumed fraction of energy released in the corona, i.e. models which are not based on a description of the accretion of corona.

The plan of the paper is the following. In Section 2 we give arguments in favour of the coronal accretion model and we explain the advantages of using the $\alpha$–viscosity description of accretion; we discuss the physical aspect of the disc/corona transition (Section 2.1) and we formulate a set of algebraic equations describing the accretion flow in the two–layer approximation (Section 2.2–5). In Section 3 we discuss the solutions obtained for the LVM. Global viscosity model computations are presented in Section 4. Our results are discussed in Section 5.

Preliminary results of this paper were presented at the IAU Symposium 159 "Active Galactic Nuclei across the electromagnetic spectrum" (Czerny 1994).

## 2 LOCAL VISCOSITY MODEL: ASSUMPTIONS AND EQUATIONS FOR THE TWO VERTICALLY AVERAGED LAYERS

The physics of accretion is poorly understood as the microscopic mechanism of the angular momentum transfer remains unknown. Most promising, perhaps, is the viscosity provided by small scale magnetic field structure which develops in the disc at the expense of its rotational energy. However, computer simulations of this process (Balbus &

Hawley 1991, 1992) do not describe usually the saturation of the process and therefore do not give the parameters for a stationary situation.

Therefore, as in the case of stellar convection, we represent all the unknown physics by a single parameter $\alpha$ introduced by Shakura & Sunyaev (1973). This kind of parameterization is very convenient. What is more, some physical attempts to describe the global effect of viscosity can be reduced to this scaling (e.g. Tout & Pringle 1992, Canuto, Goldman & Hubickyj 1984).

This assumption was widely used to calculate the vertically averaged disc structure in AGN (e.g. Ross, Fabian & Mineshige 1993, Cannizzo & Reiff 1992, Huré et al 1994a,b; see also Frank, King & Raine 1992). It was also successfully used as a local (not vertically averaged) prescription for viscosity in cataclysmic variables (Meyer & Meyer-Hoffmeister 1981; Smak 1982; see Canizzo 1994 for recent review) as well as AGN (e.g. Lin & Shields 1986; Mineshige & Shields 1990; Shimura & Takahara 1993; Cannizzo 1992).

Models which are not vertically averaged are fairly uniform in the vertical direction. However, thermal instability of the radiation pressure dominated branch as well as the tendency to develop a complex inflow pattern (e.g. Różyczka, Bodenheimer & Bell 1994) may actually lead to significant global changes of the disc structure. Therefore, it is important to explore the possibility that the resulting accretion flow consists of two different layers in the vertical direction, and that the heat generation and the radiation pressure are reduced in the optically thick layer, thus possibly leading to thermally and secularly stable solutions.

We assume that the inner accreting layer is relatively cool and optically thick (a "disc") and that the outer accreting layer is hot and optically thin (a "corona"). For a given accretion rate, a fraction of the accretion flow and the corresponding dissipation proceeds through the corona and the remaining fraction of accretion flow with the corresponding dissipation goes through the disc. These fractions are predicted by the properties of the two layers and their interaction.

In this section we consider LVM, i.e. we assume that the processes mentioned above are parameterized by a viscosity parameter $\alpha_d$ in the disc and $\alpha_c$ in the corona. In this case the disc and the corona act almost independently although there is some exchange of angular momentum and mass between them, since the fraction of accretion proceeding through the corona and the disc is a function of the distance from the black hole.

Both layers are in hydrostatic and thermal equilibrium. Their interaction occurs through the pressure balance at the base of the corona and through the exchange of radiation flux. This last interaction is described as in HM91, HM93 and NO93; it strongly influences both layers as the disc provides the source of soft photons for the corona whilst its structure is influenced by the irradiation, the reduction of heat generation and the external pressure.

We also assume that the local rotational velocity is Keplerian both in the disc and in the corona.

The numerical results will be presented only for the radius 10 $R_{\rm Sch}$, which is fairly representative of the innermost part of the disc responsible for generating most of the energy ($R_{\rm Sch}$, the Schwarzschild radius, equals $2 R_{\rm g} = 2GM/c^2$). Approximate analytical solutions at different radii are discussed in Appendix B. For the central black hole mass we take $M = 10^8 {\rm M}_\odot$, and for the viscosity parameters in disc and corona $\alpha_{\rm d} = \alpha_{\rm c} = 0.1$. We use the dimensionless accretion rate,

$$\dot m \equiv \frac{\dot M}{\dot M_{\rm Edd}},$$

where $\dot M_{\rm Edd}$ is the critical (Eddington) accretion rate

$$\dot M_{\rm Edd} = \frac{L_{\rm Edd}}{c^2 \eta} = \frac{4\pi GM m_{\rm H}}{\sigma_{\rm T} c \eta},$$

assuming the efficiency of accretion $\eta = 1/12$, as it results from the Newtonian approximation.

### 2.1 Disc/corona transition

The transition from the cold layer to the hot Compton cooled layer is neither arbitrarily positioned nor infinitely sharp. As we see from the well studied cases of weak coronae (e.g. Shaviv & Wehrse 1986, Shimura & Takahara 1993) the inversion of the temperature arises because of the drop of the atomic cooling efficiency ($\sim$ proportional to square of the density) at low densities close to the surface whilst viscous heating (proportional to the density) is still significant. Therefore, close to the surface, the electron temperature may reach the Compton temperature. The transition from atomic cooling to Compton cooling is fairly rapid in terms of gas pressure range.

In principle, full vertical structure of the gas has to be solved in order to determine the location of the transition. However, some estimates can be obtained in the optically thin regime just by studying locally the heating/cooling balance of the gas layer under the influence of viscous heating and irradiation by the upper and lower gas layers. This method is a simple generalization, for the case of (asymptotically) two-temperature medium, of the analysis made by Krolik, McKee & Tarter (1981) and Czerny & King (1989).

As a result of such calculations (see Appendix A for details) we obtain a relation between electron and ion temperatures vs. the ionization parameter, $\Xi$, defined here, after Krolik et al. (1981), as

$$\Xi \equiv \frac{F}{cP_{\rm gas}}. \tag{1}$$

where $F$ is the radiation flux. The change of gas temperature is relatively rapid in terms of pressure range (see Fig. 1). The solutions along the branch having a negative slope are thermally unstable so there should be a switch between the lower (cool) and upper (hot) branches. But there is no unique way to predict the position of this switch without computing the structure of the layer and taking self-consistently into account the flux varying with the pressure (i.e. optical depth). However, identifying this position with the upper bend should at least qualitatively well represent the properties of the transition layer.

Therefore we may define the base of the corona (i.e. the surface of the disc) by the requirement that

$$\Xi = \Xi_{\rm min}. \qquad \text{model (a)}$$

It determines uniquely the density at the disc surface if the temperature of the surface and the irradiating flux are known. We will take this condition as a component of our basic model.

**Figure 1.** Relation between the parameter $\Xi \equiv F/cP_{\rm gas}$ and electron (solid curve) and ion (dotted curve) temperatures of the plasma. The upper bend on the curve corresponds to $\Xi = \Xi_{\rm min}$, and determines the conditions at the disc – corona transition in our model (a). See Appendix A for details of calculation.

However one could choose other criteria to define the disc/corona interface, and we shall retain two of those, which also define uniquely the disc/corona system.

As our second choice we use the statement done by many authors (e.g. NO93) that the disc surface density, $\rho_{\rm s}$, is equal to the disc mean density, $\rho_{\rm d}$

$$\rho_{\rm s} = \rho_{\rm d}. \qquad\qquad {\rm model\ (b)}$$

In the light of the above discussion this condition looks rather artificial.

Another possibility is based on the fact that the optically thick part of an $\alpha$-disc is thermally unstable if dominated by radiation pressure (Lightman & Eardley 1974). The presence of the corona generally reduces the radiation pressure. Therefore we may argue that an initially weak corona will be enhanced by the developing disc instability until the radiation pressure of the cold layer drops below the instability point. Such a disc would be characterized by the marginal value of $\beta$ (gas pressure to the total pressure ratio) which is a constant resulting from stability analysis,

$$\beta = 0.4 \qquad\qquad {\rm model\ (c)}$$

(e.g. Shakura & Sunyaev 1976).

## 2.2 Cool disc

The disc structure is described by the following equations:
– hydrostatic equilibrium:
$$P_{\rm up} = P_{\rm e} - \Omega_{\rm K}^2 \rho_{\rm d} H_{\rm d}^2, \qquad (2)$$
– energy generation:
$$F_{\rm d} = \frac{3}{2}\alpha_{\rm d} P_{\rm e} \Omega_{\rm K} H_{\rm d}, \qquad (3)$$

– radiative transfer:
$$F_{\rm d} = \frac{c(P_{\rm e}^{\rm rad} - P_{\rm up}^{\rm rad})}{\kappa \rho_{\rm d} H_{\rm d}}, \qquad (4)$$
– equations of state:
$$P_{\rm e} = P_{\rm e}^{\rm rad} + \frac{k}{m_{\rm H}\mu}\rho_{\rm d} T_{\rm e}, \quad T_{\rm e} = \left(3P_{\rm e}^{\rm rad}/a\right)^{1/4}, \qquad (5)$$

$$P_{\rm up} = P_{\rm up}^{\rm rad} + \frac{k}{m_{\rm H}\mu}\rho_{\rm s} T_{\rm s}, \quad T_{\rm s} = \left(3P_{\rm up}^{\rm rad}/a\right)^{1/4}, \qquad (6)$$

and the condition that the radiation flux emitted by the disc surface (i.e. at the base of the corona) is equal to the sum of the flux generated in the disc and the fraction of the flux generated in the corona and intercepted by the disc:

$$\frac{3}{4}cP_{\rm up}^{\rm rad} = F_{\rm soft} \equiv F_{\rm d} + \eta F_{\rm c}(1-a). \qquad (7)$$

Here the symbols have the following meaning: $P_{\rm up}$ and $P_{\rm e}$ are total pressures at the base of the corona and in the equatorial plane, $P_{\rm up}^{\rm rad}$ and $P_{\rm e}^{\rm rad}$ are radiation pressures at the base of the corona and in the equatorial plane, $\rho_{\rm d}$, $\rho_{\rm s}$, $H_{\rm d}$, $F_{\rm d}$ and $\alpha_{\rm d}$ are respectively the disc mean density, the surface density, the thickness, the flux generated in the disc and the viscosity parameter in the disc, whilst $F_{\rm c}$ is the flux generated in the corona and $\mu$ is mean molecular weight of the disc gas.

The opacity $\kappa$ (the Rosseland mean) as a function of density and temperature is taken from Alexander, Johnson & Rypma (1983) for $\log T < 3.8$, from Seaton et al. (1994) for $\log T > 4.0$, and it is calculated as an interpolation from the two tables in the range $3.8 < \log T < 4.0$. We neglect the non-equilibrium influence of coronal X-rays on the opacity. This is justified as X-rays are thermalised close to the disc surface at a Thomson depth of the order of unity. The set of equations have to be supplemented with the values of $\eta$ (the fraction of the coronal flux directed towards the disc) and the X-ray albedo $a$; we assume $\eta = 0.5$ and $a = 0.15$ after HM91.

Equation (2) reduces to the standard equation of hydrostatic equilibrium in the disc if the left hand side is negligible. It can be neglected if the corona is relatively unimportant but for corona-dominated solution the pressure gradient within the disc is small as the pressure is mostly provided by the weight of the corona and one has $P_{\rm e} \approx P_{\rm up}$. This second limit corresponds to the assumption made by NO93 that the pressure in the disc interior is equal to the pressure of the corona; however, in this limit their equation (2) is not valid as it is derived under assumption that the surface pressure in the disc is negligible.

Our equation (3) reflecting the local effect of viscosity is the same as the corresponding equation of NO93 although they claim that the accretion proceeds only in the disc. SZ94 adopted different assumptions. A model based on an approach similar to their (i.e. Global Viscosity Model) is discussed in Section 4.

## 2.3 Hot corona

We assume that the hot corona is a two–temperature plasma. A physical description of such a plasma is given by Shapiro, Lightman & Eardley (1976). The gas density in the corona is taken as constant and the mean pressure in the corona is equal to the pressure at its base.

Such a corona is described by the following equations:
– hydrostatic equilibrium:

$$P_{\text{down}}^{\text{gas}} = \rho_c \Omega_K^2 H_c^2 f_{H/r} - \tau_c \frac{F_{\text{soft}}}{c}, \qquad (8)$$

where the last term represents the pressure exerted on the corona gas by the soft flux emitted by the disc,
– energy generation:

$$F_c = \frac{3}{2} \alpha_c P_{\text{down}}^{\text{gas}} \Omega_K H_c, \qquad (9)$$

– Compton cooling of electrons:

$$F_c = (e^y - 1) F_{\text{soft}}, \qquad (10)$$

– equation of state:

$$P_{\text{down}}^{\text{gas}} = \frac{k}{m_H} \rho_c \left( \frac{T_i}{\mu_i} + \frac{T_e}{\mu_e} \right), \qquad (11)$$

– total corona pressure:

$$P_{\text{down}} = P_{\text{down}}^{\text{gas}} + \frac{F_{\text{tot}}}{c}, \qquad (12)$$

where the second term on r.h.s. gives the radiation pressure on the corona side. It may be written also as $[F_{\text{soft}} + \eta(1+a)F_c]/c$.

In equation (10) we replace the amplification factor used e.g. by NO93, $y/(1-y)$, by a term which is not singular at $y = 1$, but this replacement does not actually influence any of the conclusions.

The Compton parameter $y$ for an optically thin medium is given by the formula:

$$y = \kappa_{\text{es}} \rho_c H_c \frac{4kT_e}{m_e c^2} \left( 1 + \frac{4kT_e}{m_e c^2} \right). \qquad (13)$$

Gravitational energy is converted into kinetic energy of protons which cool by Coulomb interactions with electrons,

$$F_c = \frac{3}{2} \frac{k}{m_H} \nu_{\text{ei}} H_c \rho_c (T_i - T_e) \left[ 1 + \left( \frac{4kT_e}{m_e c^2} \right)^{1/2} \right]; \qquad (14)$$

where

$$\nu_{\text{ei}} = 2.44 \times 10^{21} \ln \Lambda \, \rho_c T_e^{-1.5}; \qquad \ln \Lambda \sim 20. \qquad (14a)$$

The symbols have the following meaning: $P_{\text{down}}^{\text{gas}}$ is the gas pressure at the corona base, $\rho_c$, $H_c$, $T_i$, $T_e$ are respectively the density, the thickness of the corona, the ion and the electron temperatures, $\mu_i$ and $\mu_e$ are the mean molecular weight for ions and electrons.
Equation (8) contains the term $f_{H/r}$ which corrects for possible large geometrical thickness of the corona. It is obtained by integrating over $z$ the usual hydrostatic equilibrium equation:

$$\frac{dP}{dz} = -\frac{GM}{r^2 + z^2} \frac{z}{\sqrt{r^2 + z^2}} \rho_c, \qquad (15)$$

An approximate form of the corrective term, vaild if $H_d \ll H_c$, is

$$f_{H/r} = \left( \frac{r}{H_c} \right)^2 \left\{ 1 - \left[ 1 + \left( \frac{H_c}{r} \right)^2 \right]^{-1/2} \right\}. \qquad (16)$$

In our computations we used the exact formula obtained by integrating eq. 15 from $H_d$ to $H_c$.

This correction reproduces correctly the behaviour of the mean gravitational potential as a function of the corona thickness. The potential increases for small values of $H_c/r$, it reaches a maximum at $H_c/r \sim 1$ and decreases for $H_c/r \gg 1$. This non-monotonic behaviour leads frequently to the existence of two solutions for the disc–corona system. However, the solution corresponding to the decreasing branch of the potential is dynamically unstable and should lead to an outflow. The existence of the maximum gives a limit for the amount of energy which can be dissipated. As the right hand side of equation (8) saturates for large values of $H_c/r$ and the corona is (by definition) optically thin ($\tau_c = \kappa_{\text{es}} \rho_c H_c < 1$), the energy dissipated in the corona cannot be greater than

$$F_c^{\max} = \frac{3}{2} \alpha_c \Omega_K^3 r^2 / \kappa_{\text{es}}, \qquad (17)$$

which becomes

$$\dot{m} \leq 0.07 \left( \frac{\alpha}{0.1} \right) \left( \frac{r}{10 R_{\text{Sch}}} \right)^{1/2}, \qquad (18)$$

using eq. 19–21 written below. Therefore, for small radii and high accretion rates, the corona cannot dissipate an arbitrarily large fraction of the total energy. It is necessary to stress, however, that this behaviour is based on the coronal accretion model and that the limit depends on the value of the viscosity parameter $\alpha_c$. At the same time it is a strong prediction of the model.

### 2.4 Relation to global disc parameters

The structure of the corona depends on the total generated flux, $F_d + F_c$, and on the local Keplerian angular velocity, $\Omega_K$. Both values can be determined at a given radius $r$ from the mass of the black hole $M$ and the accretion rate $\dot{M}$ through the usual relations:

$$\Omega_K = \left( \frac{GM}{r^3} \right)^{1/2} \qquad (19)$$

and

$$F_d + F_c = \frac{3GM\dot{M}}{8\pi r^3} f(r) \qquad (20)$$

where $f(r)$ represents the boundary condition at the marginally stable orbit

$$f(r) = 1 - (3R_{\text{Sch}}/r)^{1/2} \qquad (21)$$

in the Newtonian approximation.

The disc structure in the two–layer approximation can therefore be calculated for given values of the global parameters $M$, $r$ and $\dot{M}$, and of the viscosity coefficients $\alpha_d$ and $\alpha_c$.

### 2.5 Models parameterized by the fraction of energy released in the corona

In order to better understand the properties which are tightly linked to the questionable $\alpha$–viscosity parameterization of the corona, we also calculate models which are not based on this assumption. These models are parameterized by the ratio $\xi = F_d/(F_c + F_d)$ ($\xi$ is simply related to the parameter $f$ used by HM91, HM93: $\xi = 1 - f$; it is more convenient if $f \sim 1$). They are computed without using equation (9).

## 3 LOCAL VISCOSITY MODEL: RESULTS

The most appropriate way to analyze the interaction of the corona and the disc is to solve separately the equations for the two layers, parameterizing them with $\xi$. Both sets of equations give the pressure at the base of the corona. The value of $\xi$ is then determined by equaling the pressure predicted by the disc equations to that predicted by the corona equations, $P_{\rm up} = P_{\rm down}$.

### 3.1 Qualitative discussion

In model (a) we can obtain an exact expression for the disc pressure, $P_{\rm up}(\xi)$ by combining eqs. 1, 6 and 7

$$P_{\rm up}(\xi) = \left\{ \xi \left[ \frac{4}{3} - \frac{4}{3}\eta(1-a) - \frac{\eta}{\Xi} \right] + \eta(1-a) + \frac{\eta}{\Xi} \right\} \frac{F_{\rm c} + F_{\rm d}}{c}, \quad (22)$$

so the values of $P_{\rm up}$ and of its derivative, $dP_{\rm up}/d\xi$, depend both only on global parameters. Whether or not there are pressure equilibrium solutions depends on the corona pressure. One may obtain approximate expressions for it in the two limiting cases of low $\xi$ and $\xi$ close to 1 (see below). In these two limits one can also estimate the disc pressure in model (b). Model (c) is more difficult to discuss as it represents an intermediate case between the domination of gas and radiative pressure in the disc which cannot be approximated analytically.

#### 3.1.1 Weak corona limit

In the case of a weak corona ($\xi \sim 1$) we can simplify the disc equations by assuming that the pressure at the base of the corona, $P_{\rm up}$, is much lower than the pressure in the equatorial plane, $P_{\rm e}$, that the disc is dominated by radiation pressure and that electron scattering is the only source of opacity.

Under such assumptions the relations between the pressure, the disc thickness, the density and the surface density in the disc dissipating a fraction $\xi$ of the total flux, and in the disc without a corona (i.e. $\xi = 1$; the corresponding quantities are marked with SS) are given by:

$$P_{\rm e} = P_{\rm SS}, \quad (23)$$

$$H_{\rm d} = \xi H_{\rm SS}, \quad (24)$$

$$\rho_{\rm d} = \xi^{-2} \rho_{\rm SS}, \quad (25)$$

$$\Sigma_{\rm d} = \xi^{-1} \Sigma_{\rm SS}, \quad (26)$$

i.e. the disc becomes denser and thinner as the fraction of energy released in the disc drops. At a certain point gas pressure starts to dominate. Higher density also causes the bound-free and bound-bound transitions to become the dominant source of the opacity so eq. (23)–(26) are no longer valid.

In model (b) the pressure at the top of the disc, $P_{\rm up}$, can be derived from eq. (6) by assuming that radiation pressure dominates even at the surface:

$$P_{\rm up} = \{\eta(1-a) + \xi[1 - \eta(1-a)]\} \frac{F_{\rm d} + F_{\rm c}}{c}, \quad (27)$$

so it rises with $\xi$. The values of $P_{\rm up}$ at $\xi = 1$ and of its derivative, $dP_{\rm up}/d\xi$, are therefore both linearly proportional to the accretion rate.

The pressure at the base of the corona calculated from the corona equations (8–14), $P_{\rm down}$, goes to zero for $\xi$ approaching 1. The equations can be solved analytically in this limit but they are complex owing to the difference between the electron and ion temperatures. However, combining eqs. (8) and (9), introducing the optical depth of the corona, $\tau_{\rm c} = \kappa_{\rm es}\rho_{\rm c}H_{\rm c}$ and putting $f_{H/r} = 1$, we can obtain an expression for the pressure:

$$P_{\rm down} = \left[ \frac{\tau_{\rm c}\Omega_{\rm K}(F_{\rm d} + F_{\rm c})}{\frac{3}{2}\alpha_{\rm c}\kappa_{\rm es}} \right]^{1/2} (1-\xi)^{1/2} \quad (28)$$

which shows clearly the tendency.

The condition $P_{\rm up} = P_{\rm down}$ (eq. 27 and 28) leads to proportionality

$$1 - \xi \sim \dot{M}\alpha_{\rm c}, \quad (29)$$

i.e. with increasing $\dot{M}$ the corona dissipates an increasing fraction of energy until $\xi$ becomes small.

#### 3.1.2 Strong corona limit

A detailed analytical computation of the corona dominated solutions in model (b) is carried out in Appendix B. Here we perform a qualitative discussion in order to understand more clearly the behaviour of solutions. If $\xi \ll 1$ we can simplify the equations by assuming that the disc is dominated by gas pressure and is almost isothermal, i.e. the difference between the temperatures at the base of the corona and in the equatorial plane is small and of the same order as the difference between the pressures (a more precise condition for an isothermal disc is $\xi\tau_{\rm d} \ll 1$, see Appendix B). We reach the following expressions showing trends with $\xi$:

$$P_{\rm e} = P_{\rm up} = {\rm const}, \quad (30)$$

$$T_{\rm e} = {\rm const}, \quad (31)$$

$$H_{\rm d} = \xi P_{\rm e}\kappa\Omega_{\rm K}^2, \quad (32)$$

$$P_{\rm e} - P_{\rm up} = \xi^2(P_{\rm e}\kappa\Omega_{\rm K})^2. \quad (33)$$

The exact value of the $P_{\rm up}$ obtained from this set of equations is given by:

$$P_{\rm up} = \left[ \frac{(F_{\rm d} + F_{\rm c})\Omega_{\rm K}}{\frac{3}{2}\alpha_{\rm d}\kappa} \right]^{1/2} \left( \frac{1 - T_{\rm s}/T_{\rm e}}{1 - \rho_{\rm s}T_{\rm s}/\rho_{\rm d}T_{\rm e}} \right)^{1/2}. \quad (34)$$

In model (b) the second term on the right equals 1. In model (a) the surface pressure is given by equation (22). Therefore in the first case the value of the pressure depends significantly on the opacity whilst in the second case it is roughly independent of $\xi$ for strong coronae.

The expression derived from the coronal set of equations looks similar in the limit of small $\xi$ and small luminosities (in comparison with the limit given by equation 17)

$$P_{\rm down} = \left[ \frac{\tau_{\rm c}\Omega_{\rm K}(F_{\rm d} + F_{\rm c})}{\frac{3}{2}\alpha_{\rm c}\kappa_{\rm es}} \right]^{1/2}. \quad (35)$$

Thus in model (b) the two functions $P_{\rm up}(\xi)$ and $P_{\rm down}(\xi)$ vary initially almost parallel to each other and then diverge unless the opacity varies significantly; if the initial value $P_{\rm down}(\xi \ll 1)$ is higher then $P_{\rm up}(\xi \ll 1)$, there will be a crossing point, i.e. an equilibrium corona dominated solution.

This is the case for small accretion rates as then $\kappa > \kappa_{\rm es}$ whilst $\tau_{\rm c}$ is very low. However, for high accretion rates the correction $f_{H/r}$ (see eq. 8 and 16) to the hydrostatic equilibrium becomes increasingly important and $P_{\rm down}$ decreases. Above a certain value of $\dot M$ the pressure $P_{\rm down}(\xi \ll 1)$ is smaller than $P_{\rm up}(\xi \ll 1)$ and there is no solution for the disc–corona system in pressure balance.

### 3.2 Numerical solutions for the pressures in the disc and in the corona

Before we show the results of the above relations $P_{\rm down}(\xi)$ and $P_{\rm up}(\xi)$ we demonstrate in Fig. 2 the importance of two effects included in our corona equations (8)–(14): the correction for the geometrical thickness of the corona, $f_{H/r}$, and the radiation pressure term, $\tau_{\rm c} F_{\rm soft}/c$ (eq. 8). If neither of the effects is included, the solution for $P_{\rm down}$ (parameterized by $\xi$) exists for all accretions rates and $P_{\rm down}$ is a monotonic function of $\dot m$. However, for $\dot m \sim 0.1$ the corona becomes geometrically thick ($H_{\rm c}/r > 1$; see also NO93), even if it dissipates only 10% of the gravitational energy. This shows clearly the necessity of taking into account the $H_{\rm c}/r$ correction. Its inclusion leads to a non-monotonic behaviour of $P_{\rm down}(\dot m)$: as discussed above, the pressure in optically thin corona cannot be arbitrarily large. After a maximum near $\dot m \sim 0.03$, depending on $\xi$, the pressure begins to drop. This results in an *increase* of $H_{\rm c}$ as the corona must dissipate a still increasing amount of energy according to the $\alpha$–prescription (eq. 9). Consequently, the corona also becomes optically thick contrary to the initial assumptions (and $P_{\rm down}$ increases again), but the solutions still exist formally for all $\dot m$. However, when we take into account the radiation pressure acting on the corona with $\tau_{\rm c} \sim 1$ the solutions cease to exist above $\dot m = 0.1$–0.2 (see also Appendix B); this means the corona cannot be in hydrostatic equilibrium and a possible outflow of the coronal gas must be considered (Czerny et al., in preparation)

At high accretion rates, advection constitutes an important way of transporting energy (Narayan & Yi 1994; Abramowicz et al. 1995). Rough approximation for the advective flux at radius $r$ is:

$$F_{\rm adv} = \frac{\rho_{\rm c} H_{\rm c}^2 \alpha_{\rm c} c_{\rm s}^3}{r^2}. \qquad (36)$$

Assuming that the corona dissipates bulk of the accretion power, we can solve the system of equations (8)–(14) independently of the parameters of the disc (cf. Appendix B). This leads to the result:

$$\frac{F_{\rm adv}}{F_{\rm visc}} \sim (\alpha/0.1)^{-7/6} f_{H/r}^{-1/3} \left(\frac{\dot m}{0.03}\right)^{5/6} \left(\frac{r}{10 R_{\rm Sch}}\right)^{21/12}. \qquad (37)$$

This equation shows that the advective flux should be taken into account for $\dot m$ larger than a few percent.

#### 3.2.1 Pressure in the corona

The relation $P_{\rm down}$ vs. $\xi$ for the corona is shown in Fig. 3 for three values of $\dot m$. The general behaviour is as expected ($P_{\rm down} \sim (1 - \xi)^{1/2}$; see eq. 28). For high accretion rates and lower $\xi$ the correction $f_{H/r}$ is increasingly important as $H_{\rm c}$ becomes larger than $r$ and the pressure drops because of the decrease of the gravitational potential for an enormously expanded corona. The significance of the effect is again demonstrated in Fig. 5 which shows the pressure of the corona calculated with $f_{H/r} = 1$.

**Figure 2.** Corona pressure, $P_{\rm down}$, optical depth, $\tau_{\rm c}$, and the $H_{\rm c}/r$ ratio as functions of $\dot m$ for three values of $\xi$: $10^{-4}$ (left column), 0.5 (middle column) and 0.99 (right column). In each panel the dashed curve is the solution with neither radiation pressure in the corona nor the $H_{\rm c}/r$ correction included (cf. eqs. 8 and 16). The dotted curve shows the solution with the latter correction taken into account and the solid curve is the full solution. Note the dramatic change of topology of solutions when the corrections are included.

The inclusion of the radiation pressure term in the corona equation of hydrostatic equilibrium (eq. 8) results in the existence of two solutions for $P_{\rm down}$ (and the corresponding values of other corona parameters) for each value of $\xi$ (Fig. 2). One of the solutions contradicts our assumption that the corona should be optically thin, except for $\xi \approx 1$, so we do not show this solution in the figures.

For higher accretion rates the two solutions approach each other. Above $\dot m \approx 0.1$ they form a kind of loop in the $P_{\rm down}$ vs. $\xi$ plane, i.e. they exist over a limited range of $\xi$ near unity. The range of $\xi$ shrinks as $\dot m$ increases, shifting towards larger values which mean that the solutions only exist if the corona dissipates a small amount of energy. Finally, for $\dot m > 0.2$ no solution for a dissipating corona in hydrostatic equilibrium is possible.

#### 3.2.2 Surface pressure of the disc and solution for $\xi$ in model (a)

In model (a) the transition between the disc and the corona is determined by the ratio of hard photons pressure to gas pressure, i.e. $\Xi$, such that Comptonization dominates as a cooling mechanism. Since radiation pressure at the top of the disc depends only on the partition of energy generation, i.e. $\xi$ (eq. 7), the total disc pressure at the corona base, $P_{\rm up}(\xi)$, is determined independently of the disc structure. The corona equations provide the pressure $P_{\rm down}(\xi)$, and the pressure equilibrium condition $P_{\rm up} = P_{\rm down}$ gives the solutions valid for any disc structure, provided that the constraint $P(z = H_{\rm d}) = P_{\rm up}$ can be satisfied.

**Figure 3.** Corona pressure, $P_{\rm down}$ (solid curves), and disc surface pressure, $P_{\rm up}$ (dotted and dashed curves), for three values of $\dot m$ in the three considered models of disc–corona transition (labels *a*, *b* and *c*, see Section 2.1) and both local and global viscosity models (LVM and GVM respectively, see Section 4 for definition of GVM). The number in each panel gives the accretion rate.

The disc surface pressure, $P_{\rm up}$, calculated as a function of $\xi$ on the basis of criterion (a) is shown in Fig. 3. For low accretion rates and strong coronae the predicted pressure of the corona is much higher than the pressure in the disc and the equilibrium solution exists for $\xi$ very close to 1. However, for an increasing accretion rate the height of the corona rises and the effect of the factor $f_{H/r}$ becomes increasingly important so the pressure in the corona does not grow as rapidly as the disc pressure. For accretion rates around 0.04 (see Fig. 3) both pressures are comparable and there are actually two solutions (cf Fig. 6) for the disc–corona system in pressure equilibrium. One of the solutions is characterized by large value of $H_{\rm c}/r$ (cf. Fig. 2) and it belongs to the decreasing branch of the gravitational potential. This solution is dynamically unstable. Both solutions merge at high accretion rates.

There is no solution for higher accretion rates since the disc pressure becomes larger than the corona pressure. Indeed, the disc pressure cannot be smaller than the radiation pressure determined by $\dot m$ while the corona pressure cannot be arbitrarily large due to the drop of gravitational potential.

From the disc equations one can calculate the dependencies of the disc parameters on $\xi$. For example, in Fig. 4 we show the mean disc density, $\rho_{\rm d}$, and the surface density, $\rho_{\rm s}$, as functions of $\xi$, for an accretion rate $\dot m \simeq 0.04$. The surface density varies monotonically and its value is smaller than the mean disc density for very strong as well as for very weak corona. In the intermediate range $\rho_{\rm s}$ is larger than $\rho_{\rm d}$. It will be shown in a forthcoming paper analyzing the full vertical structure of the cold layer, that the density inversion is a frequent phenomenon in radiation pressure dominated discs (Czerny et al., in preparation). In this case, however, the reason for the inversion may be the assumed value of the surface temperature, equal to $T_{\rm eff}$. Realistic values of $T_{\rm s}$ should be larger because of irradiation by coronal hard flux and, consequently, $\rho_{\rm s}$ would be lower.

**Figure 4.** Surface, $\rho_{\rm s}$, and mean, $\rho_{\rm d}$, disc densities as functions of $\xi$ in model (a), computed for $\dot m = 0.042$ and $\Xi_{\rm min} = 0.05$. Density inversions occur in some range of $\xi$. Vertical lines indicate positions of disc–corona solutions in pressure balance for this value of $\dot m$.

*3.2.3 Surface pressure of the disc and solution for $\xi$ in model (b)*

We discuss this model in some detail as it bears most similarity with published ones. It can therefore be used to show the importance of the modifications to the physical

**Figure 5.** Corona pressure, $P_{\rm down}(\xi)$ (solid curves), and disc pressures, $P_{\rm up}(\xi)$ (dotted curves) for model (b) and the same accretion rates as in Fig. 3 (LVM) but showing the influence of $H_{\rm c}/r$ correction to corona hydrostatic equilibrium equation (eq. 8) and realistic opacities in the disc case (as compared with simple Thomson opacity). To illustrate both effects the thick curves show solutions without them while the thin curves present full solutions (the same as in Fig. 3) for comparison.

description of the disc layer introduced in this paper.

The dependence of $P_{\rm up}$ on $\xi$ calculated from the disc equations, with the surface density equal to the mean density, is strongly sensitive to the details of the description. $P_{\rm up}(\xi)$ shows peculiar "bends" due to the contribution of both gas and radiation pressures. Our basic relation $P_{\rm up}(\xi)$ for the disc calculated from eq. (2–7), taking into account the contribution of bound-free and bound-bound processes to the opacities, is shown in Fig. 3. For low $\xi$ the curve is flat, as expected (see eq. 30). For $\xi$ close to 1 and high accretion rate the curve is rising (see eq. 27). For $\xi$ close to 1 but low accretion rates the asymptotic behaviour given by eq. (27) is not appropriate as the gas pressure is dominating at the base of the corona.

To demonstrate the influence of the bound-free and bound-bound processes we also plot the $P_{\rm up}(\xi)$ relation calculated with the assumption that the opacity in the disc is provided only by electron scattering. The result is shown in Fig. 5. The shape of the curve is similar to the case with full opacities but the low $\xi$ limit of $P_{\rm up}$ is now higher by at least one order of magnitude, depending on $\dot m$. This results from the fact that in the disc interior the "real" $\kappa$ is much larger than $\kappa_{\rm es}$ in this limit (cf eq. 34). For $\xi$ near 1 and $\dot m \gtrsim 0.03$ the ratio $\kappa/\kappa_{\rm es} < 2$ therefore $P_{\rm up}$ is similar to that calculated with full opacities. The existence of solutions for the coupled disc–corona system – the crossing points in Figs. 3 and 5 – is therefore very sensitive to any bend in the pressure curve.

For low accretion rate we obtain one solution for low $\xi$, i.e. with most of the energy release proceeding in the corona. This solution has the general tendency that $\xi_{\rm sol}$ decreases with an increase of $\dot m$; it disappears entirely for accretion rates higher than $\dot m \sim 0.03$. This solution resembles basically the NO93 gas dominated branch although there are significant differences in the description of the disc structure in our case and theirs (see Sec. 2.1). The disappearance of this branch for higher accretion rates is closely related to the rapid expansion of the corona and the decrease of the mean coronal gravitational potential; if this effect is not taken into account, i.e. if $f_{H/r} = 1$ the solution exists even for $\dot m = 0.1$, with $\xi$ always of the order of $10^{-3}$ (see Fig. 5, thin dashed curve and thick solid curve).

There are two additional solutions which appear over a limited range of $\dot m$ starting from $\sim 0.01$: one corresponds in most cases to $\xi \approx 1$ (a disc with a very weak corona) and the other is corona dominated ($\xi \sim 0.1$). They appear together due to the bend in the disc pressure curve and disappear also together due to disappearance of the solutions for $P_{\rm down}$ (Fig. 3), thus creating a loop on $\xi$ vs. $\dot m$ plane.

If the bound-free and bound-bound contribution to the opacity is neglected, the solutions for the disc and corona in pressure balance exist only when $\dot m > 0.03$ (Fig. 5) as the pressure in the corona is in this case almost always lower than the pressure in the disc (see eq. 34 and 35) if we assume the same value of the viscosity coefficient in the disc and in the corona. It would help to increase $\alpha$ in the disc (or to decrease $\alpha$ in the corona). However, this would be contrary to what is usually observed in cataclysmic systems, i.e. that viscosity in the hot phase is usually higher than the viscosity in the cold phase (e.g. Smak 1984); in other types of binaries this ratio is close to unity (e.g. Honeycutt, Cannizzo & Robertson 1994). More solutions could also appear if we neglect the correction for the height of the corona.

### 3.2.4 Surface pressure of the disc and solution for $\xi$ in model (c)

In our model (c) in which we require that the gas to total pressure ratio, $\beta$, is independent of $\xi$, the curve $P_{\rm up}(\xi)$ is very steep (see Fig. 3). For small values of $\xi$ there is no disc solution in hydrostatic equilibrium since $P_{\rm e}^{\rm gas}$ and, consequently $P_{\rm e}$, are small due to the $\beta = 0.4$ condition. On the other hand, for $\xi \approx 1$ the requirement for $\beta$ implies very high values of $\rho_{\rm s}$. This is because under the usual assumption $\rho_{\rm d} = \rho_{\rm s}$ the discs are strongly radiation pressure dominated. To make the gas pressure comparable to radiation pressure the mean gas density must be very large since the surface radiation pressure is given by eq. (7).

However, for intermediate $\xi$, similar to the value for which our model (b) predicts $\beta \approx 0.4$, the pressure equilibrium condition is fulfilled and a solution is possible. The surface density for this $\xi$ is lower than the mean density.

### 3.3 Properties of the disc–corona solutions

It is interesting to study the properties of a disc-like accretion flow at a given radius in the $\log \dot m$ vs. $\log \Sigma$ plane, since the slope of the relation indicates the stability or instability of the solution (e.g. Meyer & Meyer-Hoffmeister 1981, Wandel & Liang 1991, Kusunose & Zdziarski 1994). Here we perform this study at $r = 10\,R_{\rm Sch}$ and we display the values of the total surface density ($\Sigma_{\rm d} + \Sigma_{\rm c}$) as a function of $\dot m$. In this plot (Fig. 6) we show only the results of computations done for models (a) assuming $\Xi_{\rm min} = 0.05$, since they are the most realistic (see Section 2.1). For comparison, we also show a plot for the disc without corona ($\xi=1$), otherwise calculated exactly in the same way.

**Figure 6.** Accretion rate vs. surface density relation for model (a) computed adopting $\Xi_{\rm min}=0.05$. The solutions exist only up to $\dot m \approx 0.085$. For comparison, the solid curves show the relation in models parameterized by the fraction of energy generated in the disc, $\xi$.

For small accretion rates the model with a corona follows closely the branch without corona. The fraction of energy generated in the corona is minor and it has almost no influence on the disc structure. Above the accretion rate $\sim 0.04$ multiple solutions are possible (see also Fig. 3). They form a kind of loop in this diagram. The horizontal branch results from the fact that the relations $P_{\rm down}(\xi)$ and $P_{\rm up}(\xi)$ are parallel for $\xi \lesssim 0.1$ (Fig. 3) so, when they cross, the entire range of $\xi$ is obtained within a very small range of $\dot m$. The other branch of the $\log \dot m$–$\log \Sigma$ diagram, along which the coronal accretion is important, has actually a negative slope which usually suggests some kind of instability although a more detailed discussion of stability is necessary to confirm it (see e.g. NO93).

For higher accretion rates (above $\dot m \approx 0.085$) there is no solution for the disc–corona system. The pressure in the corona is lower than the pressure required at the disc surface for any partition of the dissipation. It is related to the fact that the corona being in hydrostatic equilibrium can dissipate only a limited amount of energy (see eq. 17).

Although the solution for the disc-corona system is not unique when expressed as a function of accretion rate, the fraction of energy liberated in the disc varies monotonically along the loop. The same is true for all the other parameters except for very low accretion rate: the dependence of the temperature of the corona, of its geometrical thickness and the Compton parameter on the accretion rate, are shown in Fig. 7. The change at low accretion rate is connected with the fact that the solutions become single-temperature, i.e. the proton temperature is equal to the electron temperature is that range.

All these results depend on the assumed value of the viscosity parameter $\alpha$. A decrease of its value lowers the curves in the $\log \dot m$–$\log \Sigma$ diagram, since the maximum accretion rate for a disc-corona system with an accreting corona in hydrostatic equilibrium depends linearly on $\alpha$ (see eq. 18). Lower values of $\alpha$ also lead to a decrease of the optical depth and an increase of both the ion and electron temperatures in the corona, as we can see from the analytical discussion of Appendix B. Eq. (18) and the relations given in Appendix B also indicate how the results vary with radius. Although for larger radii the limit for the accretion rate becomes less stringent the role of advection term increases (see eq. 37).

Results for model (b) plotted on $\log \dot m$ vs. $\log \Sigma$ diagram would form a number of branches, including a long corona-dominated branch with a positive slope covering a broad range of accretion rates. Such a solution resembles the solution of NO93. Although this idea might be attractive the existence of this branch is not well justified physically; it simply reflects that fact that the surface density in case (b) is calculated as a mean density and depends on the amount of flux generated in the disc for $\xi \ll 1$, whilst in model (a) this surface density is calculated from the ionization parameter and is roughly independent on $\xi$ in this regime.

**Figure 7.** Coronal parameters for solutions of model a) shown as functions of the accretion rate, $\dot m$: $\xi$, electron and ion temperatures $T_{\rm e}$ and $T_{\rm i}$, comptonization parameter $y$, and $H_{\rm c}/r$ ratio.

Results for model (c) plotted on $\dot{m}$–$\Sigma$ diagram would be positioned along a straight vertical line. This can be most easily seen from eqs. 2–5 supplemented with $P_{\rm e}^{\rm rad}/P_{\rm e} = $ const condition. Assuming for simplicity $P_{\rm up} \ll P_{\rm e}$ and $P_{\rm up}^{\rm rad} \ll P_{\rm e}^{\rm rad}$ one obtains a solution for all disc parameters which is independent of $F_{\rm d}$ and hence of $\dot{m}$.

### 3.4 Comparison with models $\xi$ = const

The range of solutions parameterized by $\xi$ is bounded in $\dot{m}$ (Fig. 6). The upper limit reflects the impossibility of obtaining solutions in hydrostatic equilibrium even if the $\alpha$-prescription in the corona (eq. 9) is not used. There is also a lower limit coming from the requirement of a large optical thickness in the disc which is not fulfilled when $\dot{m}$ and $\xi$ are small.

Solutions parameterized by constant $\xi$ form a series of curves on the $\dot{m}$–$\Sigma$ diagram whose slope indicates stability. One sees (Fig. 6) that weak corona solutions are still unstable due to the dominance of radiation pressure (Lightman–Eardley instability) in the disc above certain $\dot{m}$. The critical value of $\dot{m}$ increases with decreasing $\xi$ and, for $\xi \lesssim 0.1$, the disc is always stable.

### 3.5 Role of pair production

At high coronal temperatures pair production may be important. The estimated value of the compactness parameter for our corona–dominated solutions is, however, moderate:

$$l_{\rm c} \equiv \frac{\sigma_{\rm T}}{m_{\rm e}c^3} \frac{H_{\rm c}}{r^2} L_{\rm C} \simeq 10^4 \frac{H_{\rm c}}{r} \frac{\dot{m}(1-\xi)}{(r/R_{\rm Sch})} \le 200. \quad (38)$$

($H_{\rm c}/r \sim 1$, $\xi \ll 1$, $\dot{m} \le 0.05$ at $r = 10\,R_{\rm Sch}$). Therefore we have solved only *a posteriori* the equation of pair balance to check how important they can be. For conditions corresponding to the disc–corona solutions (Fig. 7) we have computed the optical depth due to pairs and compared it with the total optical depth of the corona. We proceeded in a similar manner as HM93. The balance equation for thermal pairs can be written (Zdziarski 1985, HM93)

$$\tau_{\rm p}^2 = \tau_{\rm c}^2(1-\Lambda), \quad (39)$$

where $\tau_{\rm p}$ and $\tau_{\rm c}$ are the "proton" and total optical depths, respectively, (it is $\tau_{\rm c}$ that results from coronal energy balance) and

$$\Lambda = \left\{ \left(\frac{N_{\rm P}}{n_{\rm e}}\right)^2 \left[f_{\rm PP} + \frac{N_{\rm W}}{N_{\rm P}} f_{\rm PW} + \left(\frac{N_{\rm W}}{N_{\rm P}}\right)^2 f_{\rm WW} \right] \right\} / f_{\rm A}. \quad (40)$$

The input spectrum (presumably due to Comptonization) is defined as the sum of a power law and a Wien peak and

**Figure 8.** The corona total optical depth, $\tau_{\rm c}$, resulting from coronal energy balance (filled squares), the "proton" optical depth, $\tau_{\rm p}$ (open circles) and the optical depth due to pairs, $\tau_{\rm ee}$, (stars) computed for the solutions of LVM model a) (cf Fig. 6 & 7)

$N_{\rm P}$ and $N_{\rm W}$ are constants multiplying the two parts. Their ratio can be expressed as (Zdziarski 1985)

$$\frac{N_{\rm W}}{N_{\rm P}} = \frac{\Gamma(\alpha)}{\Gamma(2\alpha+3)} P_{\tau_{\rm c}}, \quad (41)$$

where $P_{\tau_{\rm c}}$ is the mean probability of scattering. We calculate $\alpha$ and $P_{\tau_{\rm c}}$ from formulae given by Zdziarski et al. 1994, appropriate for optically thin uniform slab (although we realize that $H_{\rm c}/r \sim 1$ means rather conical geometry). The terms $f_{\rm PP}$, $f_{\rm PW}$ and $f_{\rm WW}$ are pair production rates due to photon–photon interactions (we neglect any other pair production mechanisms) while $f_{\rm A}$ is the annihilation rate. Pair escape is neglected in our calculations. For all the rates we employ expressions given in HM93 (note that the formula B3 for $f_{\rm A}$ in that work is misprinted), Svensson (1984) and Zdziarski (1985). Finally, the input spectrum is normalized to the density of Comptonized photons which can be estimated (HM93) as

$$N_\gamma \simeq \frac{L_{\rm C}(1+\tau_{\rm c})}{c \langle E_{\rm C} \rangle} \frac{1}{2\pi r(r+H_{\rm c})}, \quad (42)$$

where $L_{\rm C}$ is the energy released in the corona, $\langle E_{\rm C} \rangle$ being the mean photon energy.

Fig. 8 displays the values of the three optical depths $\tau_{\rm c}$, $\tau_{\rm p}$ and the optical depth due to pairs, $\tau_{\rm ee}$, for solutions of our model a). Under the most favourable conditions $\tau_{\rm ee}$ does not exceed 15% of $\tau_{\rm c}$, i.e. the pairs are relatively unimportant. The reason for this is a combination of non-relativistic temperatures ($\Theta \equiv kT_{\rm e}/m_{\rm e}c^2 \le 0.4$) and significant $\tau_{\rm c} \sim 0.5$. Under these circumstances the compactness parameter required for efficient pair production is $\ge 1000$ (see, e.g., Fig. 3 in HM93).

## 4 GLOBAL VISCOSITY MODEL

### 4.1 Description and equations

As we actually do not know how the viscosity mechanism operates we should allow for a range of possibilities. If the viscosity is connected with generation of magnetic field we can either assume that the corona itself generates its own local magnetic field – and therefore creates a local viscosity mechanism – or assume that the cool disc is the only source of magnetic field. In this latter case the magnetic field of the disc would simply penetrate into the corona thus possibly extracting angular momentum from the hot gas and transporting this angular momentum to the disc. The hot corona, loosing its angular momentum, can accrete; the cool disc through its own local viscosity mechanism, has to transport the entire angular momentum whilst participating only partially to the accretion flow and to the corresponding dissipation.

Such a picture can be translated into equations in a very simple way using the description of disc and corona structure given in Sections 2.2-3. We only require that the entire angular momentum (not a fraction) is transported by the cool disc. This requirement, for a Keplerian disc, is formally equivalent to replacing $F_d$ in eq. (3) with $F_d + F_c$ as it was done by SZ94, leaving the fraction of the flux to be transported by radiation (eq. 4) unchanged.

We cannot (in principle) describe now the corona using the equation (9) since the local viscosity do not operate there any more. Instead, we can assume that the radial velocity is a constant fraction of the sound velocity. If we call this fraction $\alpha_c$ and assume that the corona is in hydrostatic equilibrium then equation (9) has to be replaced by:

$$\xi \dot{M} = 4\pi r \alpha_c v_s \rho_c H_c; \qquad v_s = \left(\frac{P_{\text{down}}}{\rho_c}\right)^{1/2}. \tag{43}$$

For a geometrically thick corona ($H_c \sim r$ or larger) this new equation is in practice equivalent to the previous one, the only difference being the factor $f(r)$ representing the inner boundary condition for an accretion disc (eq. 21).

The predictions for the disc structure from the GVM differ strongly from the predictions based on the viscosity mechanism operating locally. The equations of the cold layer structure (Section 2.2) lead to a different asymptotic behaviour if the dissipation of energy in the disc is small. Instead of a trend qualitatively represented in the LVM by eq. 23 – 26 we find in this case:

$$P_e = P_{\text{up}} = \left[\frac{(F_d + F_c)\Omega_K}{\frac{3}{2}\alpha_d \kappa}\right]^{1/2} \frac{1}{\xi^{1/2}}, \tag{44}$$

$$H_d \sim \xi^{1/2} \tag{45}$$

so the pressure diverges for small dissipation inside the disc. It simply results from the fact that the disc has to transport all the angular momentum whilst dissipating none. The structure of the corona predicted on the basis of the limit for the radial velocity is not significantly different from LVM.

### 4.2 Results

The numerical computations confirm the above described trend (Fig. 3). In model (b) the pressure expected at the base of the corona is much higher than for a standard disc. The surface pressure in model (a) is defined independently of the disc structure so it is the same as in LVM (cf Fig. 3), while in model (c) $P_{\text{up}}(\xi)$ differs from that in LVM but not significantly. The corona pressure, $P_{\text{down}}$, is also considerably modified (Fig 3, note that the panels for LVM and GVM do not show the same accretion rates). For a given accretion rate it is lower than in LVM and the solutions exist only up to $\dot{m} \approx 0.07$, in comparison with $\dot{m} \approx 0.2$ in LVM, for the adopted values of parameters.

**Figure 9.** Accretion rate vs. surface density relation for global viscosity model a) (circles). The curve shows LVM for comparison.

Disc–corona solutions in model (a) have a character very similar to those in LVM: a weak corona is obtained for small accretion rates. For $\dot{m} \sim 0.02$ (if $\Xi_{\min} = 0.05$) the asymptotic values $P_{\text{up}}(\xi \ll 1)$ and $P_{\text{down}}(\xi \ll 1)$ are comparable and a corona dominated branch of solution occurs. In model (b) the solutions exist only in a narrow range of $\dot{m} \gtrsim 0.02$ and they are disc-dominated.

On the $\dot{m}$–$\Sigma$ diagram (Fig. 9) model (a) solutions form two branches of negative slopes. The horizontal branch extends towards larger $\Sigma$, instead of smaller $\Sigma$ as in LVM. So in GVM there is no turn characteristic for $\alpha$–disc solutions occurring when the gas and radiation pressures are comparable. Model (c) solutions are positioned along a straight line of positive slope but the averaged quantities (density, pressure) are much larger in the global model than in the local one. Density inversions do not occur in this model.

The opacity is high in GVM. Consequently the temperature gradient is large even if the disc dissipates only a tiny fraction of the energy because of the increase in the total optical depth by four orders of magnitude with respect to a standard disc which dissipates the same fraction of energy but transports only the correspondingly small fraction of angular momentum. The mass of such a disc becomes considerable if the disc extends to more than 100 $R_{\text{Sch}}$ and it becomes self-gravitating.

## 5 DISCUSSION AND CONCLUSIONS

In this paper we made an attempt to describe the disc-corona system without specifying arbitrarily the partition of the energy generation between the two layers, but instead predicting it. We considered two cases. In the first one the corona as well as the disc is accreting and transporting the angular momentum through a viscosity parameterized by $\alpha$ (LVM – local viscosity model). In the second one we assumed that the entire angular momentum is transported vertically to the disc and radially through the disc (GVM – global viscosity model). The partition between the disc and the corona in our favourite model (a) (see Section 2) was determined by the critical value of the ionization parameter $\Xi$. In both cases we assumed that the corona can be described as a continuous medium.

The direct conclusions from this study are the following:

i) at this stage we cannot favour any of the two models of angular momentum transport. Both predict stable discs for accretion rates below 0.001 of the Eddington limit (when gas pressure dominates) and unstable discs between 0.001 and 0.1 (with the presence of a relatively weak corona). There is a corona dominated (perhaps) stable solution in the LVM model but for a very narrow range of accretion rates.

ii) the present modeling of the corona is not suitable for higher accretion rates because of the limit on the flux generated in the corona if it is in hydrostatic equilibrium. An outflow from the corona in the vertical direction is expected. This effect, together with radial advection, may change the picture for high accretion rates,

iii) when the parameters of the disc–corona system are calculated as functions of the fraction of energy liberated in the disc, they depends (but not critically) on the underlying assumptions,

iv) for final solutions in pressure balance between the disc and the corona the fraction of energy liberated in the disc depends strongly on details of the description: the adopted law of the disc–corona transition, description of opacities within the disc, the correction for corona geometrical thickness,

v) our physically justified solutions do not possess a stable corona–dominated solution of the type found by NO93, covering a broad range of accretion rates because such a solution (obtained in case (b) – see Section 2) is based on the assumption that the surface density is equal to the mean density in the disc. The mean density varies with the fraction of energy generated in the disc $\xi$ even if it is low (i.e. $\xi \ll 1$) whilst the surface density calculated from the ionization parameter $\Xi$ does not depend on $\xi$ in this limit.

The present study did not give a full picture of accretion so it is not possible at this stage to discuss the possible advantages of a model of continuous corona over a clumpy corona recently proposed by Haardt, Maraschi & Ghisellini (1994).

Future investigations should therefore go into two parallel directions. First, the theoretical approach has to solve the structure of the disc–corona system for high accretion rates, and to predict its dependence on the values of the parameters. Second, the observational approach should try to give an insight into the relation between the amplitudes of variations of the UV flux (which in the disc model would correspond to disc thermal instabilities, e.g. Siemiginowska & Czerny 1989), the relative level of X–ray and UV emission, and the Eddington ratio.

If the dynamical study of the disc–corona system confirms our present result that there is no solution for large Eddington ratios, one should expect strong differences in the UV-X spectrum of objects accreting close to or far from the Eddington rate. Many studies of the broad line emission of AGN have shown that quasars and Seyfert galaxies differ not only in their luminosity, but also in their Eddington ratio, close to unity for quasars, and at least one order of magnitude smaller in Seyfert galaxies. An important observational issue would therefore be to clarify the link between the luminosity of AGN and their other properties. Preliminary results seem to indicate that the so-called "reflection" component observed in the X-ray spectrum of Seyfert galaxies and due to reprocessing of X-rays on cold matter, is absent in quasars (Williams et al. 1992). In Seyfert galaxies the existence of reprocessing between UV and X-ray photons is also ascertained by the absence of time lag between them (Clavel et al. 1991). Moreover it is well-known that the X-ray to UV flux ratio is much smaller in quasars than in Seyfert galaxies. All these facts could indicate that there is indeed a fundamental difference between the accretion process in quasars and in Seyfert galaxies, and the presence or absence of a strong accreting corona above the disc could be a clue for this problem.

## ACKNOWLEDGMENTS

This project was partially supported by PICS/CNRS no. 198 "Astronomie Pologne", by French Ministry of Research and Technology within RFR programme, and by grant no. 2P30401004 of the Polish State Committee for Scientific Research.

## APPENDIX A

We can estimate from simple algebraic equations the temperature of an optically thin layer of gas at a density $\rho$ close to that of the base of the corona. Such a layer is irradiated both from the top (by upper corona) and from the bottom (cool disc) and it also participates in the energy dissipation due to the Keplerian shear and to the presence of viscosity.

Here we assume that the viscous heating is proportional to the gas pressure of ions,

$$Q^+_{\rm visc} = \frac{3}{2}\alpha\Omega_{\rm K}\frac{k}{m_{\rm H}}T_{\rm i}. \tag{A1}$$

This energy goes primarily to ions and later on it is transfered to electrons via Coulomb interactions. The difference of temperature between electrons and protons in the stationary case is given by:

$$Q^+_{\rm visc} = \frac{3}{2}\frac{k}{m_{\rm H}}\nu_{\rm ei}\rho(T_{\rm i} - T_{\rm e}), \tag{A2}$$

(for definition of $\nu_{\rm ei}$ see eq. 14a).

Apart from viscous heating, the layer is heated by absorption of outgoing cool photons, which we describe by the Kramers opacity coefficient. Compton scattering of incoming and outgoing photons gives a net heating or cooling, depending on the sign of the difference between the electron temperature and the Compton temperature of the radiation bath. The complex atomic processes providing an efficient cooling mechanism at $10^5 - 10^6$ K can be approximated by a simple analytical formula for the cooling function $\lambda(T_{\rm e})$ (Buff & McCray 1974).

Values of $T_{\rm i}$ and $T_{\rm e}$ resulting from the energy balance can be conveniently plotted vs. $\Xi$ (see eq. 1). The plot is not very accurate as the approximation of the cooling function is not satisfactory for lower temperatures but the position of the upper bend on the curve is reasonably correct. The characteristic value of $\Xi$ corresponding to this bend depends most strongly on the adopted value of the Compton temperature. It is equal to 0.5 for $T_{\rm IC} = 4\times 10^8$ K and 0.05 for $10^9$ K. As the extension of the observed primary X-ray component of AGN spectra suggest rather high temperature, above 100 keV, we adopted the second value of $\Xi$ in most of our computations.

## APPENDIX B: ANALYTICAL SOLUTION OF THE CORONA DOMINATED CASE FOR THE LOCAL VISCOSITY MODEL (B)

To understand qualitatively the behaviour of the solutions found with numerical calculations, it is interesting to perform an analytical study, which moreover give their dependence on various parameters.

### B1  General conditions for the existence of the corona

In the corona dominated case ($\xi \ll 1$) the system of equations is simplified, since $F_{\rm c} = F_{\rm t}$, where $F_{\rm t}$ is the sum $F_{\rm d}+F_{\rm c}$. We assume the same value of $\alpha$ in the disc and in the corona.

To simplify the discussion, we limit it to model b), which corresponds to the largest range of $\dot{m}$ giving a corona dominated solution. The disc structure is then described by the following equations (to avoid confusion, we shall keep the same numeration as in the main text).

The hydrostatic equilibrium equation (2) is not modified. The energy generation equation writes:

$$\xi F_{\rm t} = \frac{3}{2}\alpha P_{\rm e}\Omega_{\rm K}H_{\rm d}, \tag{B3}$$

and the radiative transfer equation:

$$\xi F_{\rm t} = \frac{c(P^{\rm rad}_{\rm e} - P^{\rm rad}_{\rm up})}{\kappa\rho_{\rm d}H_{\rm d}}, \tag{B4}$$

The equations of state (B5) and (B6) are identical to equations (5) and (6), respectively, and equation (7) writes:

$$\frac{3}{4}cP^{\rm rad}_{\rm up} = \eta(1-a)F_{\rm t}. \tag{B7}$$

The corona is described by the following equations: hydrostatic equilibrium:

$$P_{\text{down}}^{\text{gas}} = \rho_c \Omega_K^2 H_c^2 f_{H/r} - \tau_c \frac{\eta(1-a)F_t}{c} \quad (B8)$$

energy generation:

$$F_t = \frac{3}{2}\alpha P_{\text{down}}^{\text{gas}} \Omega_K H_c \quad (B9)$$

Compton cooling of electrons:

$$F_t = (e^y - 1)\eta(1-a)F_t \quad (B10)$$

The following equations, (11), (12), (13) and (14), are not changed. Completed with equations (15), (19), (20) and (21), and with the condition $P_{\text{up}} = P_{\text{down}}$, they allow to determine all parameters as functions of $M$, $\dot{M}$ and $r$. In the following we shall use $M_8$, the mass in $10^8 \, M_\odot$, $\dot{m}_{0.03}$, the dimensionless accretion rate in $0.03 \dot{M}_{\text{Edd}}$ (both values being typical of Seyfert nuclei), $r_{10}$, the radius in $10 R_{\text{Sch}}$, $f_{10} = f(r)/f(r_{10})$ and $\alpha_{0.1} = \alpha/0.1$. Expressed in these units $F_t$ and $\Omega_K$ become:

$$F_t = 1.4 \times 10^{14} \dot{m}_{0.03} \, f_{10} \, M_8^{-1} \, r_{10}^{-3} \quad \text{erg cm}^{-2} \, \text{s}^{-1}, \quad (B19)$$

and

$$\Omega_K = 2.2 \times 10^{-5} M_8^{-1} \, r_{10}^{-3/2} \, \text{s}^{-1}. \quad (B20)$$

To simplify we shall also assume that $\eta(1-a)$ is constant and equal to 0.5 (this is true within 20%), $\mu$=0.5, $\mu_i$ and $\mu_e = 1$.

One can first show that **there is no corona dominated solution if $P_{\text{down}}$ is dominated by radiation pressure**, i.e. by the term $F_t/c$. If it is the case, and since $\tau_c$ is of the order unity, equation (B8) shows that $P_{\text{down}}^{\text{gas}}$ is dominated by the second term on the right – the dynamic radiation pressure – and is therefore negative (unless the two terms on the right balance almost exactly, which implies that $F_t$, and consequently $\dot{M}$, is very small according to equation [B9]). This solution has no physical meaning, since the corona would be bloated by radiation pressure.

Assuming then that the first term of the r.h.s. of equation (B8), $\rho_c \Omega_K^2 H_c^2 f_{H/r}$, is much larger than the second, the equations of the corona can be solved independently of $\xi$, and lead to the following solution (we assume $T_i \gg T_e$ and we neglect the second order term in [B13] and [B14])

$$\rho_c = 7.1 \times 10^{-15} \, \alpha_{0.1}^{3/4} \, f_{H/r}^{1/4} \, M_8^{-1} \, (\dot{m}_{0.03} f_{10})^{-1/4} \, r_{10}^{-9/8}, \quad (B21)$$

$$\frac{H_c}{r} = 0.77 \, \alpha_{0.1}^{-7/12} \, f_{H/r}^{-5/12} \, (\dot{m}_{0.03} f_{10})^{5/12} \, r_{10}^{-1/8}, \quad (B22)$$

$$\tau_c = 0.56 \, \alpha_{0.1}^{1/6} \, f_{H/r}^{-1/6} \, (\dot{m}_{0.03} f_{10})^{1/6} \, r_{10}^{-1/4}, \quad (B23)$$

$$T_e = 2.9 \times 10^9 \, \alpha_{0.1}^{-1/6} \, f_{H/r}^{1/6} \, (\dot{m}_{0.03} f_{10})^{-1/6} \, r_{10}^{1/4}, \quad (B24)$$

$$T_i = 3.2 \times 10^{11} \, \alpha_{0.1}^{-7/6} \, f_{H/r}^{1/6} \, (\dot{m}_{0.03} f_{10})^{5/6} \, r_{10}^{-5/4}. \quad (B25)$$

These expressions show that the corona satisfies easily three of the four required conditions: the ion gas is non relativistic ($T_i < 10^{13}$K), $H_c$ is not much larger than $r$, and $\tau_c$ is of the order unity. None of these conditions depend on the mass of the black hole. The fourth condition, that the electron gas be also non relativistic ($kT_e < m_e c^2$, i.e. $T_e < 6 \times 10^9$ K) is more stringent, and in particular it implies that the radius of the corona is limited to about $100 R_{\text{Sch}}$ with this type of solution. A consistent computation should take into account the corrective terms in equations (13) and (14). However, since the compactness parameter drops with $r$ much faster than $T_e$ increases (see eq. 38) the importance of electron–positron pairs should not be significant at large distances.

Equation (B22) shows that **the non dimensional accretion rate is limited to $\dot{m}$ less than 0.03, in order for the corona to be geometrically thin**. This limit is the same as that given by the dominance of the advective flux on the viscous flux in the corona.

Finally the pressure of the corona is given by:

$$P_{\text{down}} = 1.8 \times 10^5 \, \alpha 1^{-5/12} \, f_{H/r}^{5/12} \, M_8^{-1} \times \\ (\dot{m}_{0.03} f_{10})^{7/12} \, r_{10}^{-19/8}. \quad (B26)$$

One should now check if there is a solution for the disc, with $P_{\text{up}} = P_{\text{down}}$. It is clear that if $P_{\text{up}} = P_{\text{down}}$ and if gas pressure dominates in the corona, $P_{\text{up}}$ is also dominated by gas pressure. This is due to the fact that the isotropic radiation pressure at the disc surface and the dynamical pressure at the base of the corona are of the same order, so when it dominates in the corona, it also dominates in the disc. We are therefore led to the conclusion that **gas pressure dominates $P_{\text{up}}$**, if a solution exists.

Since $P_{\text{up}}$ should be dominated by gas pressure, one gets from equations (B6) and (B7):

$$\rho_d = 2.35 \times 10^{-8} \, \alpha_{0.1}^{-5/12} \, f_{H/r}^{5/12} \, M_8^{-3/4} \times \\ (\dot{m}_{0.03} f_{10})^{1/3} \, r_{10}^{-13/8}. \quad (B27)$$

The condition $P_{\text{up}}^{\text{gas}} \gg P_{\text{up}}^{\text{rad}}$ writes then:

$$32 \, \alpha_{0.1}^{-5/2} \, f_{H/r}^{5/12} (\dot{m}_{0.03} f_{10})^{-5/12} \, r_{10}^{5/8} \gg 1. \quad (B28)$$

Assuming $\alpha_{0.1}$, $f_{H/r}$ and $f_{10}$ equal to unity, it leads finally to:

$$\dot{m} \ll 0.12 \, r_{10}^{3/2}, \quad (B29)$$

which does not give a limit on the accretion rate as stringent as the advective flux.

### B2  Solutions for the disc

Equations (B4) and (B7) give:

$$P_e^{\text{rad}} = \frac{F_t}{c} \left( \xi \tau_d + \frac{2}{3} \right). \quad (B30)$$

where $\tau_d$ is the optical thickness of the disc.

The second term on the r.h.s. is $\approx P_{\text{up}}^{\text{rad}}$, and it dominates if $\xi \tau_d \ll \frac{2}{3}$. This is the general condition for radiative heating to dominate viscous heating **inside** the disc (cf. Huré et al. 1994ab). Let us examine the two cases.

### B2.1  $\xi \tau_d \gg \frac{2}{3}$ or $P_e^{\text{rad}} \gg P_{\text{up}}^{\text{rad}}$.

We are again facing two possibilities. Either $P_e$ is also dominated by gas pressure, or it is dominated by radiation pressure.

*B2.1.1   $P_e$ is dominated by gas pressure.* Then:

$$P_e = \frac{k\rho_d}{\mu m_H} \left(\frac{3\xi F_t \tau_d}{ac}\right)^{1/4}. \tag{B31}$$

Since $P_e$ and $P_{up}$ are both dominated by gas pressure, and $P_e^{rad} \gg P_{up}^{rad}$, one deduces that $P_e \gg P_{up}$, and equation (C2) becomes simply:

$$P_e = \rho_d H_d^2 \Omega_K^2. \tag{B2bis}$$

From the set of equations (B2bis), (B3) to (B7), (B30) and (B31), one gets the solution for the disc:

$$\rho_d = 7 \times 10^{-7} B^{-3/10} \xi^{2/5} \alpha_{0.1}^{-7/10} M_8^{-7/10} \times (\dot{m}_{0.03} f_{10})^{2/5} r_{10}^{-33/20}, \tag{B32}$$

$$\xi = 1.6 \times 10^{-4} B^{3/4} \alpha_{0.1}^{17/24} f_{H/r}^{25/24} M_8^{-1/8} \times (\dot{m}_{0.03} f_{10})^{-1/6} r_{10}^{1/16}, \tag{B33}$$

$$\frac{H_d}{r} = 3.4 \times 10^{-4} B^{1/4} \alpha_{0.1}^{1/24} M_8^{-1/8} (\dot{m}_{0.03} f_{10})^{1/6} r_{10}^{1/16}, \tag{B34}$$

$$\tau_d = 10^3 B^{5/4} \alpha_{0.1}^{-3/4} f_{H/r}^{5/12} M_8^{1/8} (\dot{m}_{0.03} f_{10})^{1/2} r_{10}^{-9/16}, \tag{B35}$$

where $B \equiv \kappa/\kappa_{es}$ is generally of the order of 10.

An important condition is that the disc stays optically thick, in order to radiate a quasi blackbody continuum in the visible-UV range. With this solution, we see that the disc is geometrically thin and optically thick.

The value of $\xi$ is consistent with our basic assumption $\xi \ll 1$. It depends very little on the value of the physical parameters, and it is of the order of $10^{-3}$. Its exact value can be determined by a numerical computation giving $B$.

Computing the value of $P_{up}$ from equations (6), (B27) and (B32), one gets:

$$P_{up} = 3 \times 10^6 \, \xi^{2/5} B^{-3/10} \alpha_{0.1}^{-7/10} M_8^{-7/10} (\dot{m}_{0.03} f_{10})^{-2/5} r_{10}^{-33/20}, \tag{B36}$$

which shows that $P_{up}$ increases with $\xi$.

Now we have to check the assumptions made to establish this solution.

The condition $\xi \tau_d \gg \frac{2}{3}$ gives:

$$\frac{2}{3} \ll 0.16 \, B^2 \, \alpha_{0.1}^{-1/24} f_{H/r}^{35/24} \, (\dot{m}_{0.03} f_{10})^{1/3} \, r_{10}^{-1/2}. \tag{B37}$$

For $B = 10$, and assuming again $\alpha_{0.1} = f_{H/r} = f_{10} = 1$, it leads to a condition on the accretion rate:

$$\dot{m} \gg 2 \, 10^{-6} \, r_{10}^{1/2}, \tag{B37bis}$$

which is easily satisfied at any radius.

The condition that $P_e$ is dominated by gas pressure writes:

$$150 \, B^{-3/2} \, \alpha_{0.1}^{-2/3} \, f_{H/r}^{-5/6} \, (\dot{m}_{0.03} f_{10})^{-2/3} \, r_{10}^{1/3} \gg 1, \tag{B38}$$

which gives a condition on the accretion rate (again $\alpha_{0.1}$, $f_{H/r}$ and $f_{10} =1$):

$$\dot{m} \ll 60 \, r_{10}^{1/2} \, B^{-9/4}. \tag{B38bis}$$

This condition is strongly dependent on the value of $B$, but even for $B \sim 10$, it gives reasonable accretion rates $\dot{m} \ll 0.3 r_{10}^{1/2}$ for any value of the radius.

**In conclusion the solution exists at all radii, for any accretion rate $\dot{m}$ smaller than 0.03. It corresponds to $\xi \sim 10^{-3}$ and to $P_{up}$ increasing with $\xi$.**

*B2.1.2   $P_e$ is dominated by radiation pressure.* Then

$$P_e = \frac{\xi F_t \tau_d}{c}. \tag{B39}$$

Equation (C2bis) is still valid, since again $P_e$ is $\gg P_{up}$. From the set of equations (C2bis), (C3) to (C8), and (C39), one gets the solution for the disc:

$$\rho_d = 1.5 \times 10^{-9} B^{-3} \xi^{-2} \alpha_{0.1}^{-1} M_8^{-1} (\dot{m}_{0.03} f_{10})^{-2} r_{10}^{3/2}, \tag{B40}$$

$$\xi = 0.25 \, B^{-3/2} \alpha_{0.1}^{-7/24} f_{H/r}^{-5/24} M_8^{-1/8} (\dot{m}_{0.03} f_{10})^{-7/6} r_{10}^{25/16}, \tag{B41}$$

$$\frac{H_d}{r} = 1.6 \times 10^{-3} B^{-1/2} \alpha_{0.1}^{-7/24} f_{H/r}^{-5/24} M_8^{-1/8} (\dot{m}_{0.03} f_{10})^{-1/6} r_{10}^{9/16}, \tag{B42}$$

$$\tau_d = 4.5 \times 10^3 \, B^{1/2} \alpha_{0.1}^{-17/24} f_{H/r}^{5/24} M_8^{1/8} (\dot{m}_{0.03} f_{10})^{1/6} r_{10}^{-1/16}. \tag{B43}$$

The value of $\xi$ is consistent with our basic assumption $\xi \ll 1$ and the disc is geometrically thin and optically thick. But $\xi$ is larger than in the previous solution, since it is $\sim 10^{-2}$ and, contrary to the previous case, it depends strongly on the radius. So the solution should disappear for $r_{10}$ larger than a few units. From equations (6), (B27) and (B40), we get:

$$P_{up} = 6.4 \times 10^5 \, \xi^{-2} B^{-3} \alpha_{0.1}^{-1} M_8^{-1} (\dot{m}_{0.03} f_{10})^{-2} r_{10}^{-3/2}, \tag{B44}$$

which shows that $P_{up}$ decreases now with $\xi$.

The condition $\xi \tau_d \gg \frac{2}{3}$, gives:

$$\frac{2}{3} \ll 1.1 \, 10^3 \, B^{-1} \, \alpha_{0.1}^{-1} \, (\dot{m}_{0.03} f_{10})^{-1} \, r_{10}^{3/2}, \tag{B45}$$

For $B = 10$, $\alpha_{0.1} = f_{H/r} = f_{10} = 1$, it leads to a condition on the accretion rate:

$$\dot{m} \ll 6 \, r_{10}^{3/2}, \tag{B45bis}$$

which is satisfied at any radius.

The condition that $P_e$ is dominated by radiation pressure writes:

$$40 \gg B \, \alpha_{0.1}^{7/12} \, f_{H/r}^{5/12} \, (\dot{m}_{0.03} f_{10})^{7/12} \, r_{10}^{-7/8}, \tag{B46}$$

which gives a condition on the accretion rate (with $\alpha_{0.1} = f_{H/r} = f_{10} = 1$):

$$\dot{m} \ll 0.4 \, r_{10}^{3/2} \, B^{-12/7}. \tag{B46bis}$$

This condition is again strongly dependent on the value of $B$. For $B \sim 10$, it corresponds to a quite small accretion rate, $\dot{m} \ll 10^{-2} r_{10}^{3/2}$. However the limit on the accretion rate increases rapidly with the radius.

In conclusion the solution exists at all radii for small accretion rates ($\dot{m} \ll 10^{-2}$). It should disappear for $r_{10}$ larger than a few units because $\xi$ becomes too large. The solution corresponds to $\xi \sim 10^{-2}$ for $r_{10} \sim 1$ and to $P_{\rm up}$ decreasing with $\xi$.

### B2.2 $\quad \xi \tau_{\rm d} \ll \frac{2}{3}$ or $P_{\rm e}^{\rm rad} = P_{\rm up}^{\rm rad} = 2 F_{\rm t}/3c$.

This hypothesis implies that $P_{\rm e}^{\rm gas} = P_{\rm up}^{\rm gas}$, so **the disc is isothermal, $P_{\rm e}$ is dominated by gas pressure, and $P_{\rm e} = P_{\rm up}$**. It is the case considered by NO93. It leads however to an inconsistency, since one gets $H_{\rm d} = 0$ from equations (A1) and (A2). This is due to the fact that the density at the surface of the disc, $\rho_{\rm s}$, is actually not exactly equal to the density at the equatorial plane, $\rho_{\rm d}$. Let us call $A \equiv \rho_{\rm s}/\rho_{\rm d} < 1$. Equation (A1) then writes:

$$P_{\rm e} = \frac{\Omega_{\rm K}^2 \rho_{\rm d} H_{\rm d}^2}{1 - A}. \tag{B2ter}$$

Note that NO93 avoid the singularity by setting simply $P_{\rm e} = \Omega_{\rm K}^2 \rho_{\rm d} H_{\rm d}^2$. Their equation is an acceptable approximation to (B2ter) only if $A$ is small, and it is not consistent with the condition $P_{\rm up} > \Omega_{\rm K}^2 \rho_{\rm d} H_{\rm d}^2$.

The set of equations (B2ter), (B3) to (B7) yield the solution for the disc:

$$\xi = 2.8 \times 10^{-4} \frac{\sqrt{1-A}}{A} \alpha_{0.1}^{7/12} M_8^{-1/8} \\ (\dot{m}_{0.03} f_{10})^{-7/24} r_{10}^{1/4} f_{H/r}^{5/12}, \tag{B47}$$

$$\frac{H_{\rm d}}{r} = 3.3\, 10^{-2} \sqrt{(1-A)}\, M_8^{7/8}\, (\dot{m}_{0.03} f_{10})^{1/8}\, r_{10}^{1/8}, \tag{B48}$$

$$\tau_{\rm d} = 10^3\, B\, \sqrt{1-A}\, \alpha_{0.1}^{-5/12}\, M_8^{1/8} \\ (\dot{m}_{0.03} f_{10})^{11/24}\, r_{10}^{-1/2}\, f_{H/r}^{5/12}, \tag{B49}$$

As $\tau_{\rm d}$ must satisfy the condition $\xi \tau_{\rm d} \ll \frac{2}{3}$, then finally $\tau_{\rm d}$ should verify $1 \ll \tau_{\rm d} \ll \frac{2}{3\xi}$. The first condition is easily satisfied, but the second one gives:

$$0.36\, B\, \frac{1-A}{A}\, (\dot{m}_{0.03} f_{10})^{1/6}\, r_{10}^{-1/4} \ll \frac{2}{3}, \tag{B50}$$

which can be satisfied only if $A$ is very close to unity. We conclude that **this solution is not consistent, and the disc is never completely isothermal**.

### B3  Conclusion

To summarize, we have found that two solutions for the corona dominated case, corresponding to two different values of $\xi$, exist for small accretion rates, one being limited to small radii, and the other to $10^2$ gravitational radii. For larger radii, the corona becomes relativistic and pair effects should be taken into account.

(1) We have shown that the only case where an optically and geometrically thin corona can be maintained in hydrostatic equilibrium is when the radiative pressure at the surface of the cold disc is negligible, and it corresponds to a limit on the accretion rate: $\dot{m} \ll 0.12\, r_{10}^{3/2}$, which does not give a limit on the accretion rate as stringent as the advective flux.

(2) At "small" values of $\xi$, $\sim 10^{-3}$, there is a solution for a geometrically thin and optically thick disc with $P_{\rm e} > P_{\rm up}$, $P_{\rm e}$ being dominated by gas pressure. This solution exists at large radii and relatively large accretion rates. The completely isothermal solution with $P_{\rm e} = P_{\rm up}$, as the one proposed by NO93, does not exist.

(3) At "intermediate" values of $\xi$, $\sim 10^{-2}$, a solution exists for a geometrically thin and optically thick disc with $P_{\rm e} > P_{\rm up}$, $P_{\rm e}$ being dominated by radiation pressure. This solution is limited to small accretion rates and to small radii.

This rather approximate study leads to several conclusions which agree with those obtained from the numerical computation for the same parameters (radius, mass, viscosity parameter). Obviously it allows only to know if there is a solution of the disc-corona model in an asymptotic case, when either the radiative, or the gas pressure, is negligible. Other solutions might exist in intermediate cases when none of the pressures dominates completely, but this can be checked only by detailed computations.

This paper has been produced using the Blackwell Scientific Publications TEX macros.

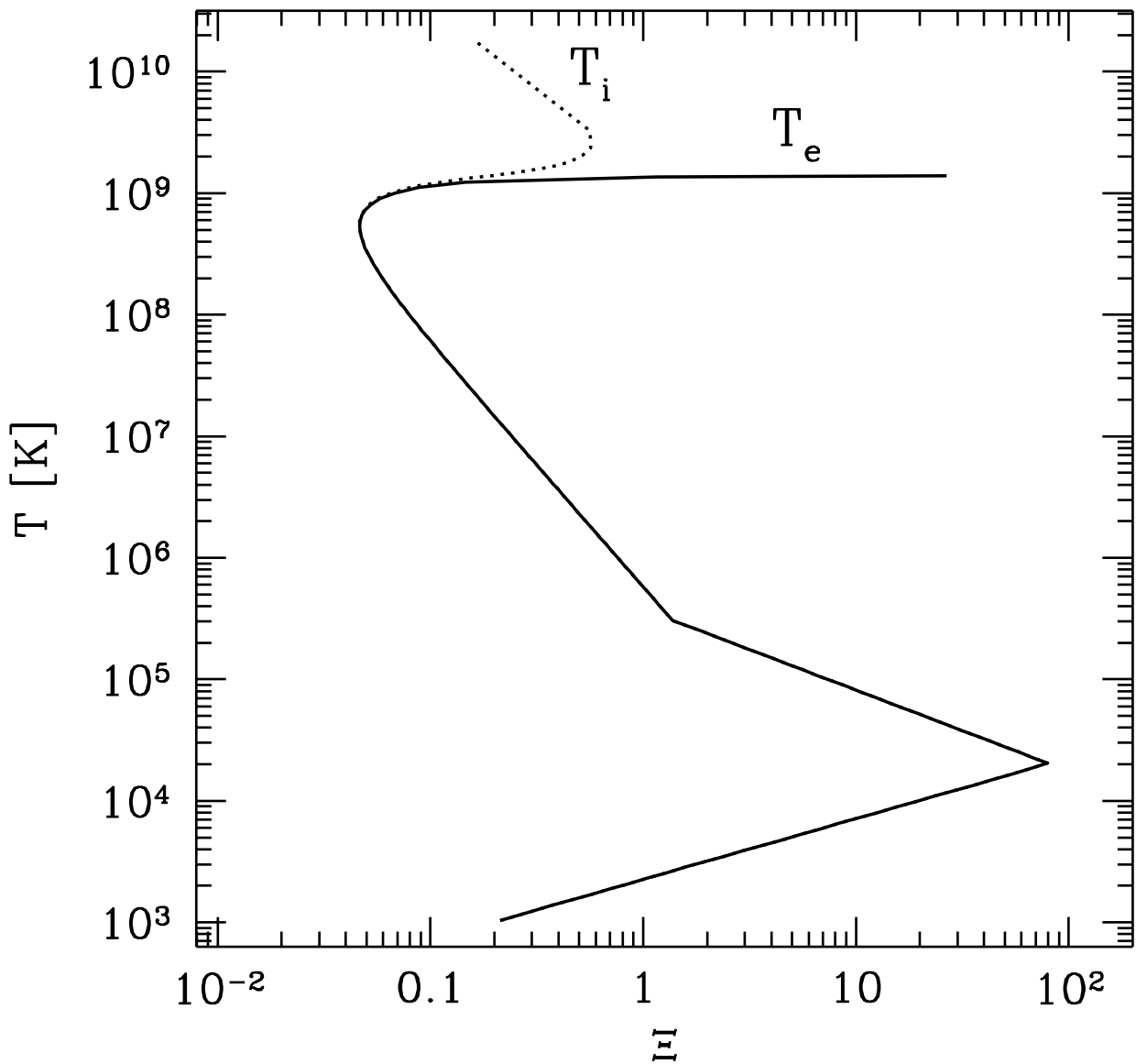

Fig. 1

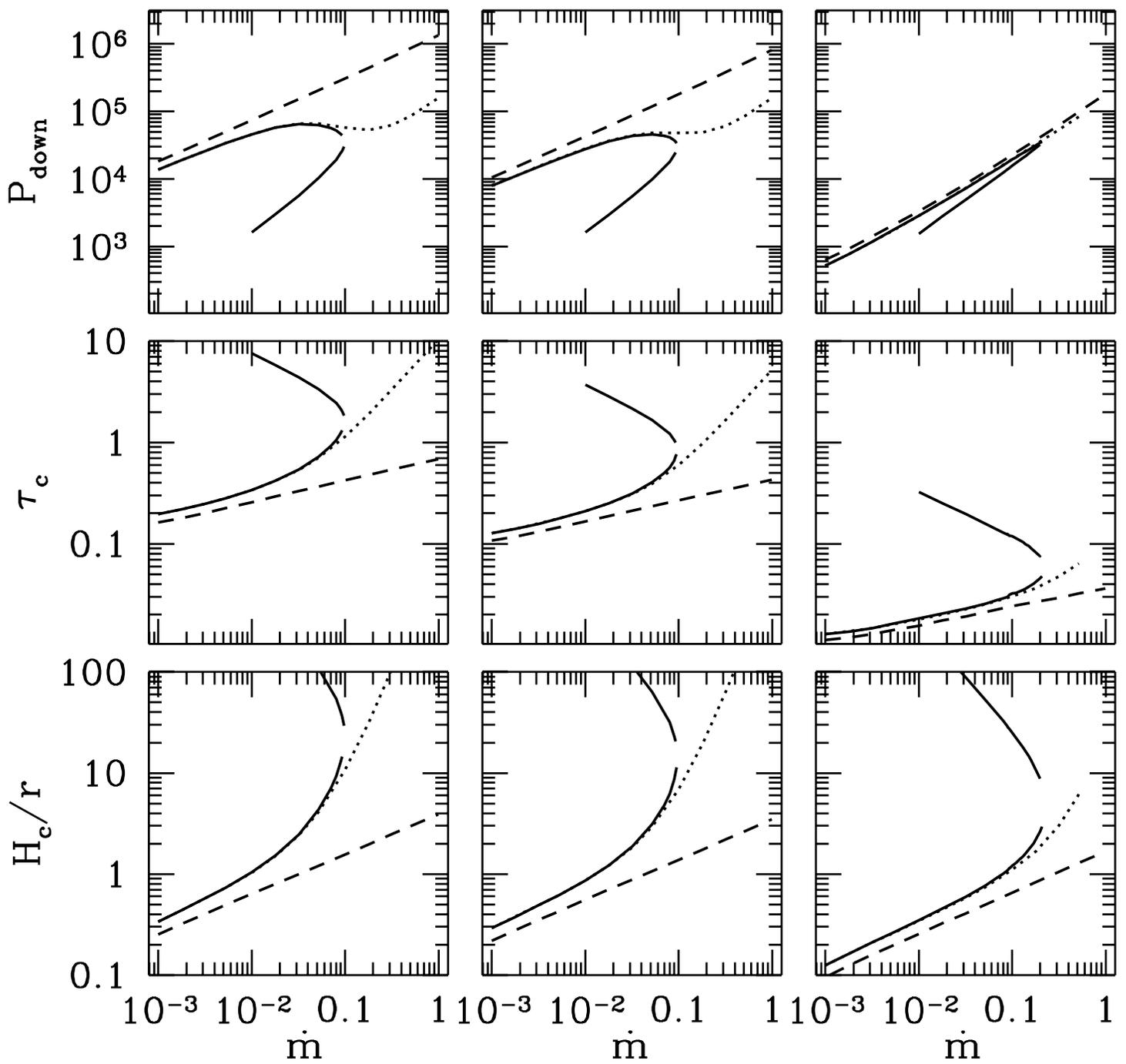

Fig. 2

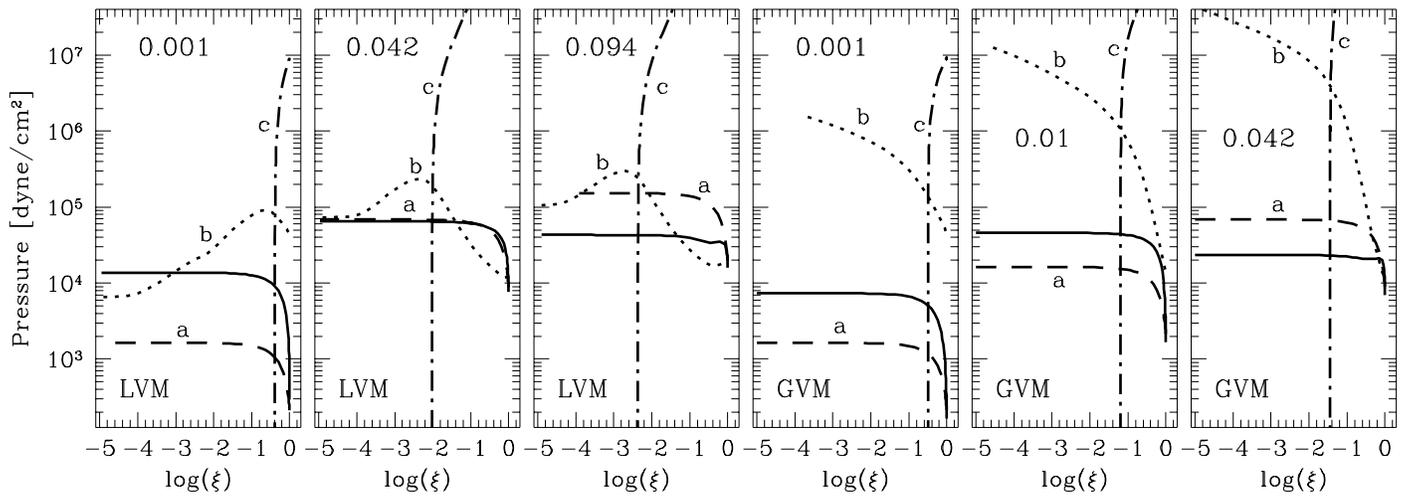

Fig. 3

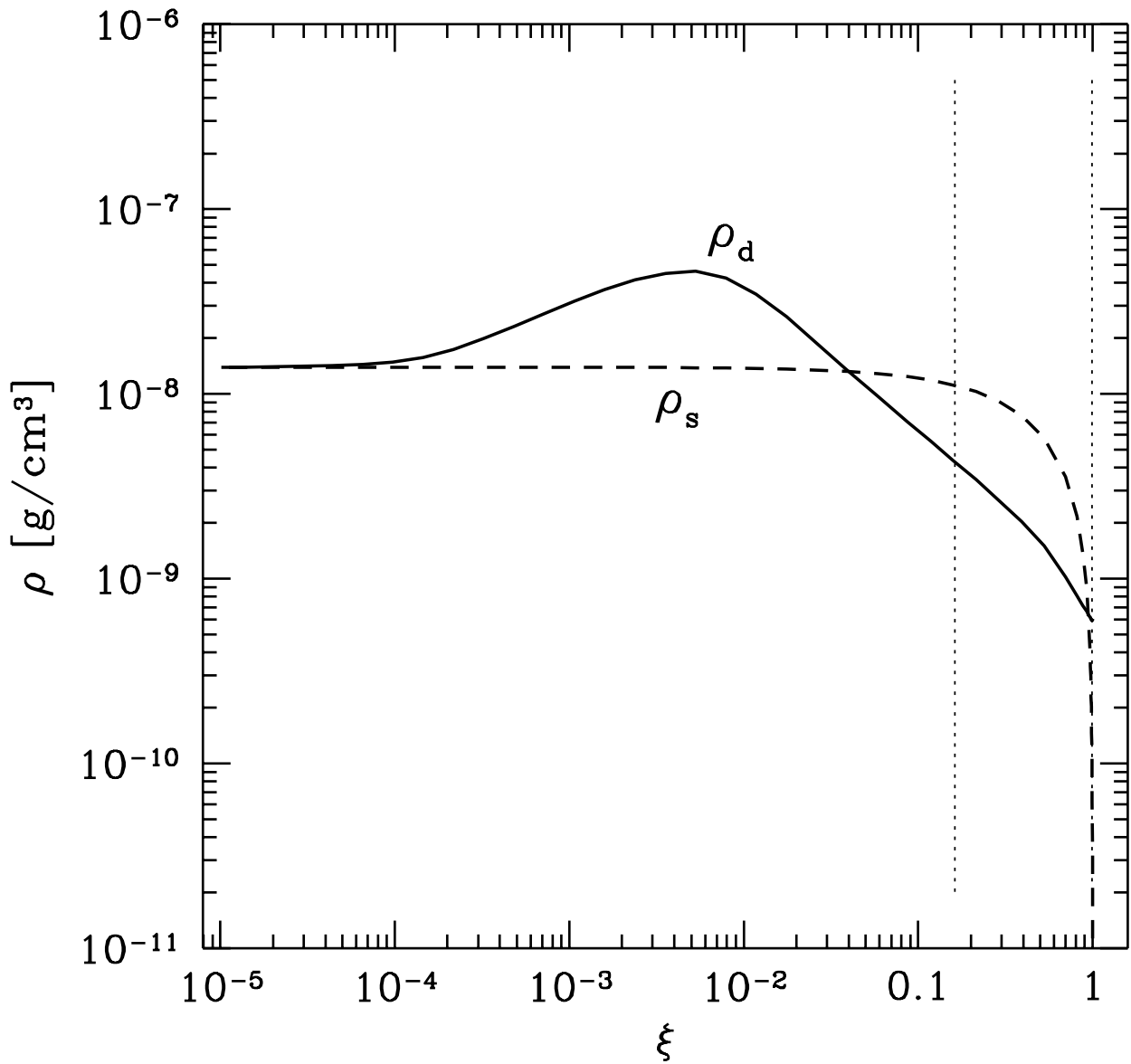

Fig. 4

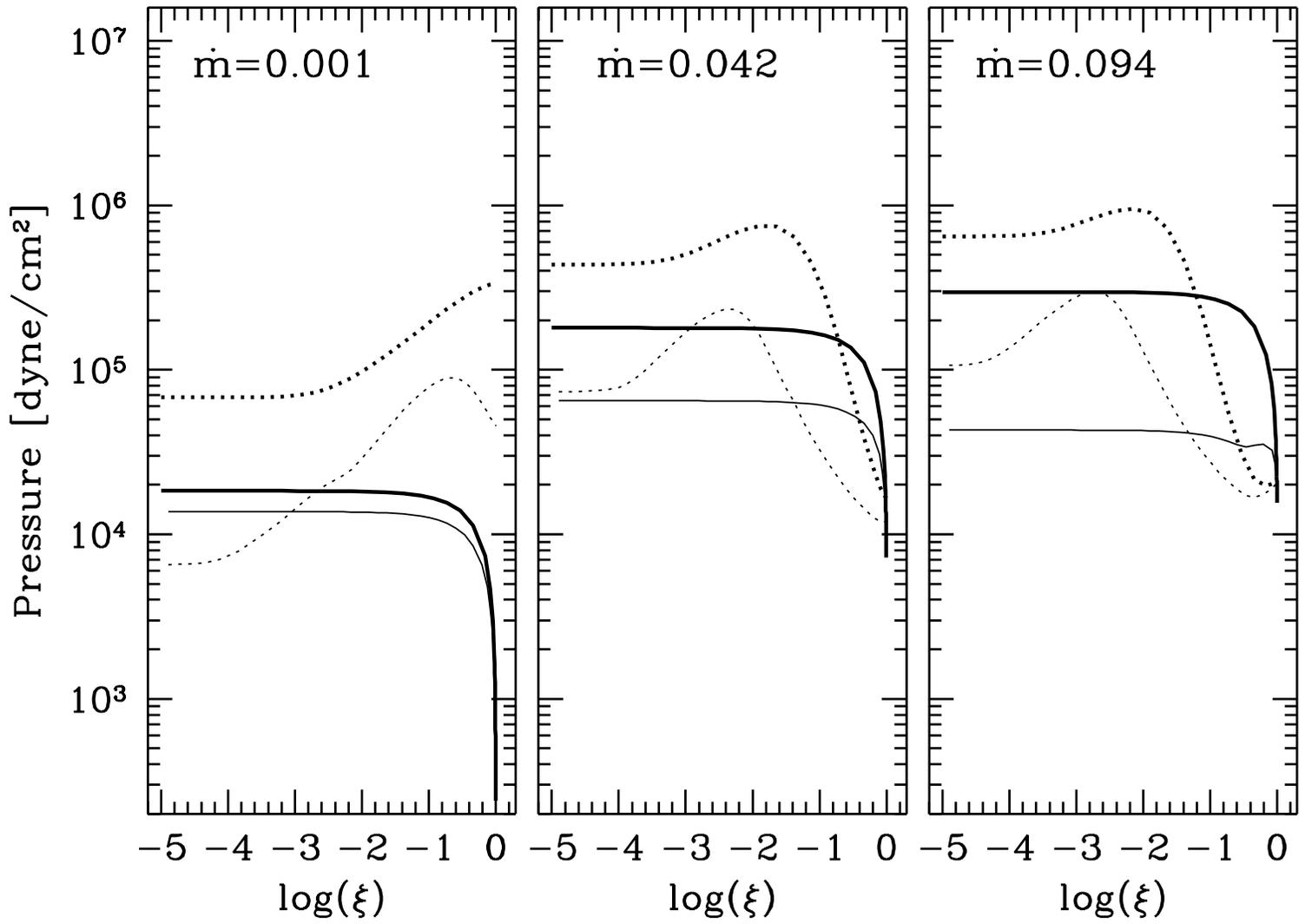

Fig. 5

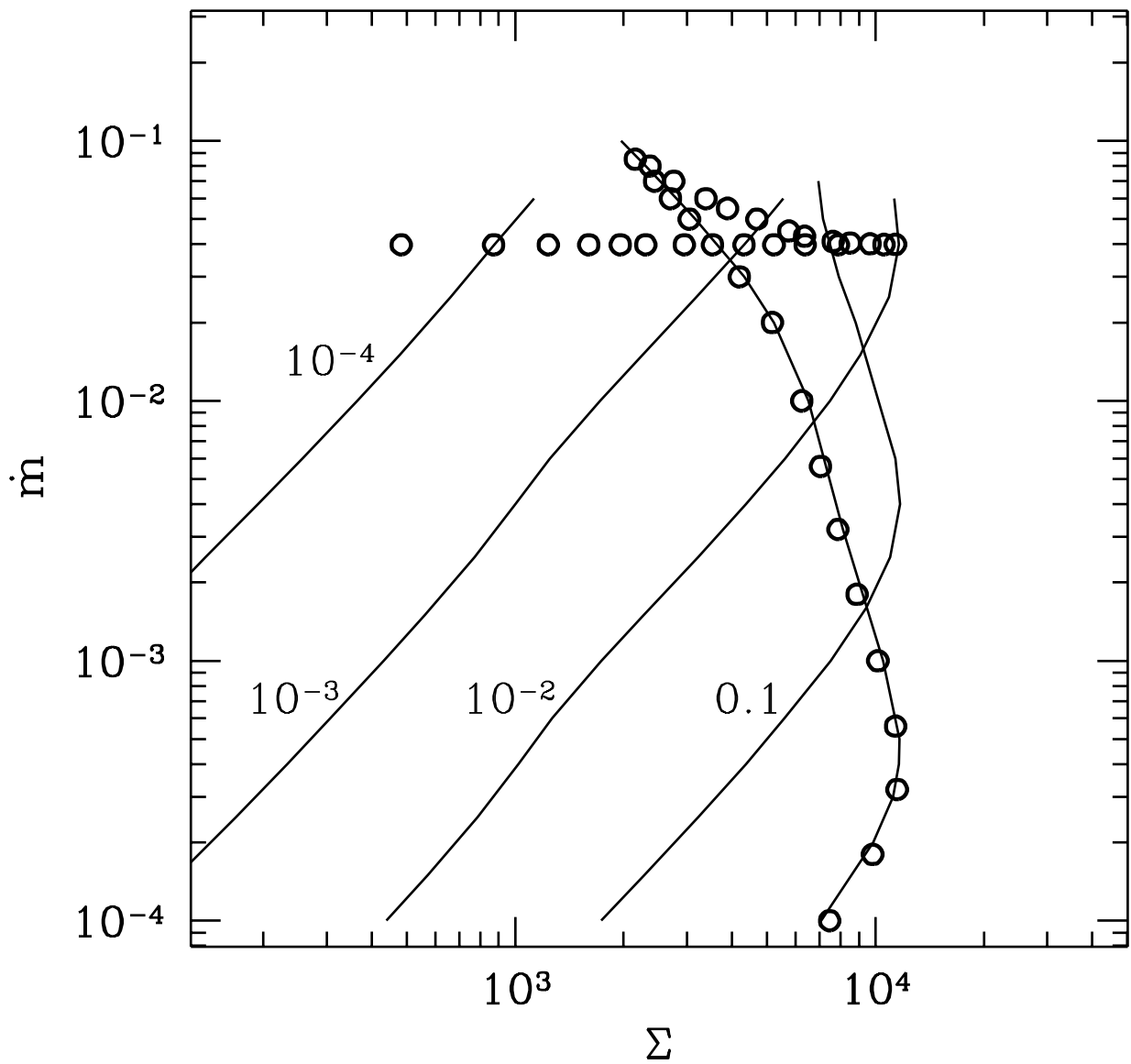

Fig. 6

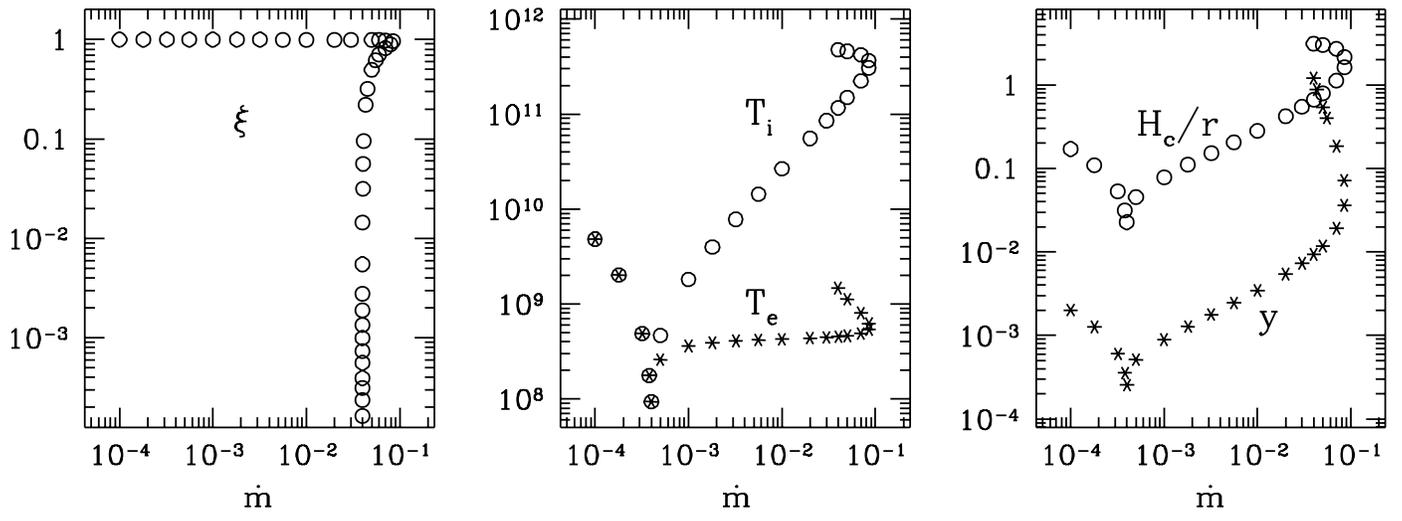

Fig. 7

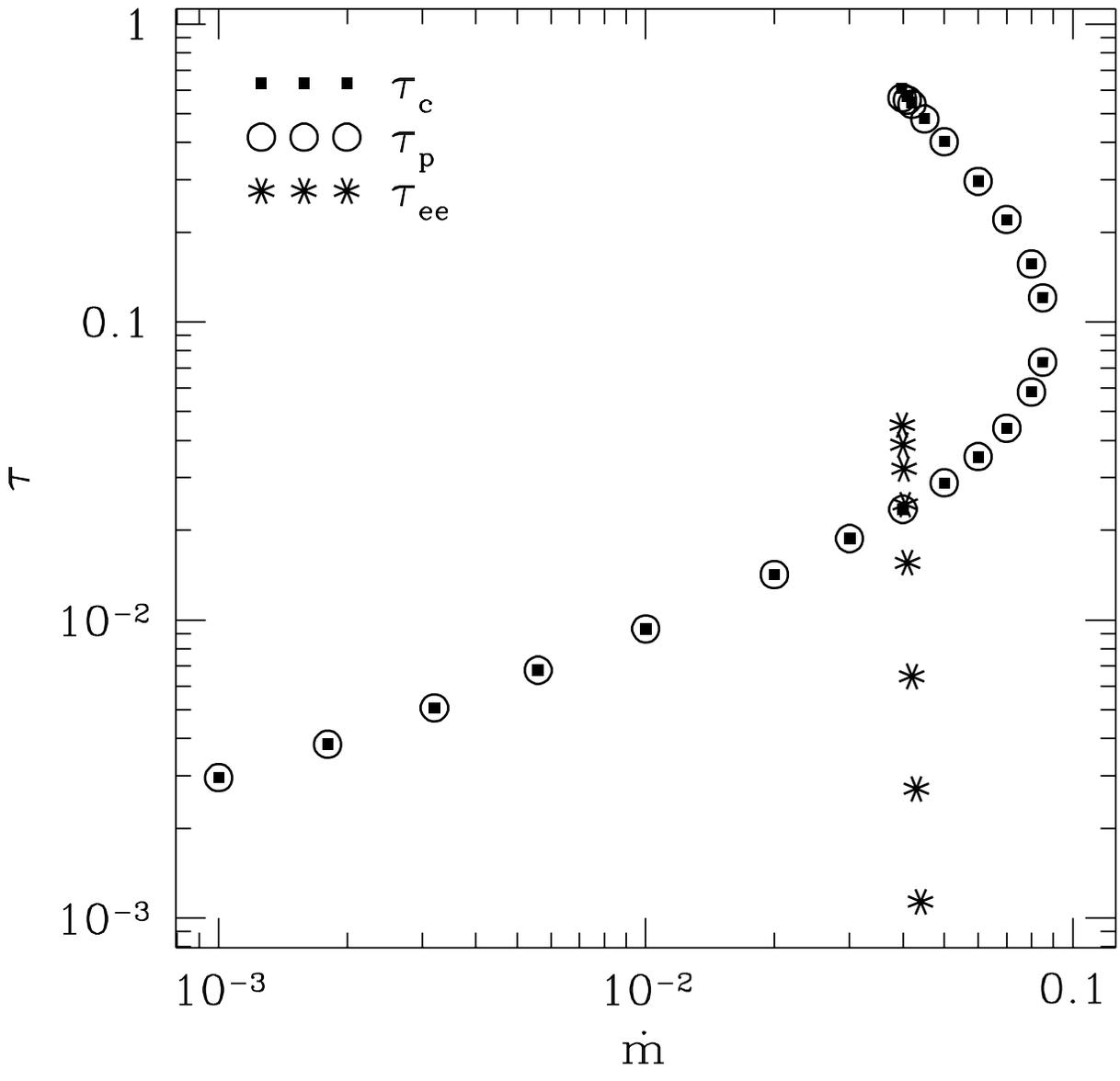

Fig. 8

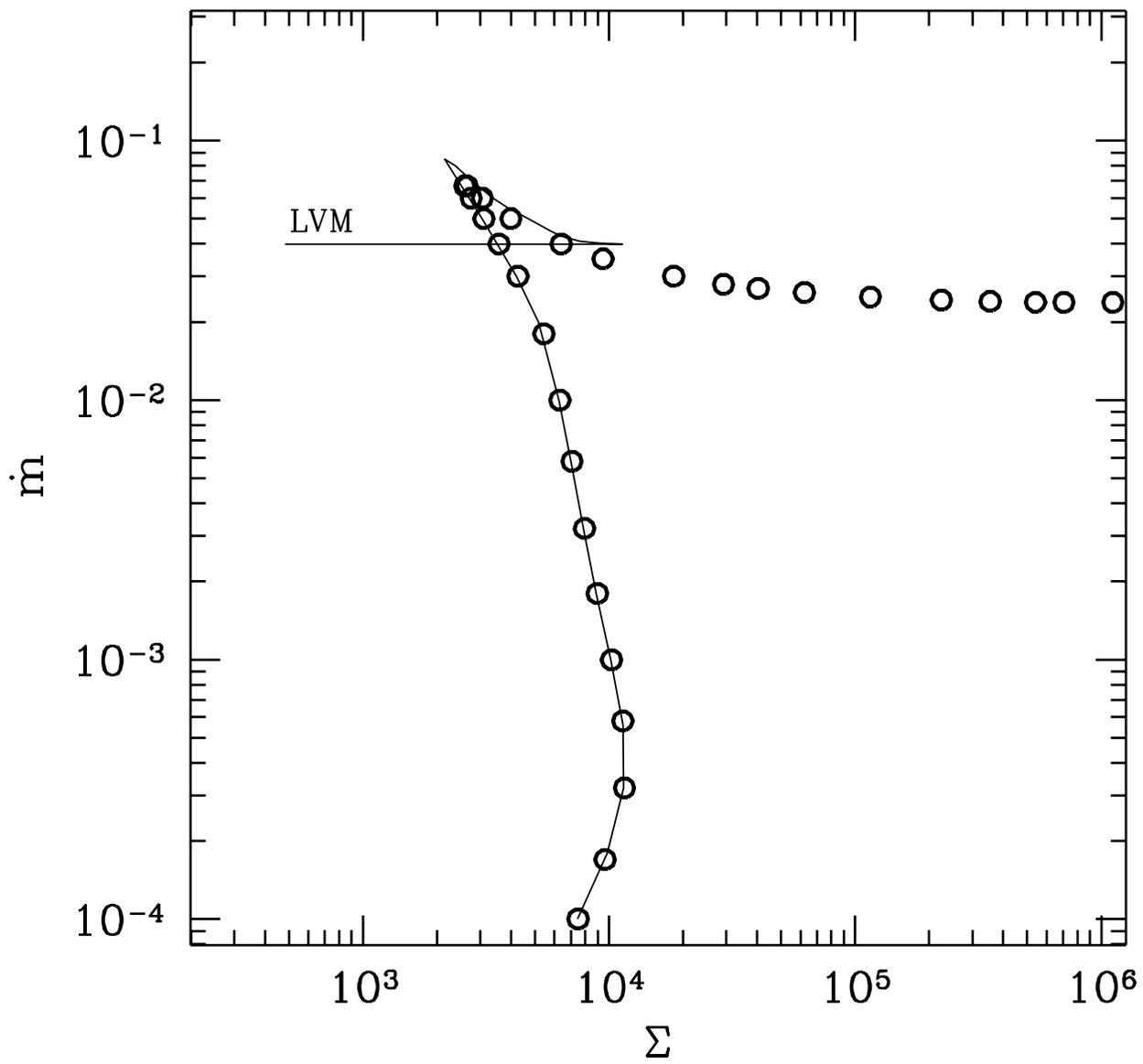

Fig. 9